\documentclass[aps,pra,twocolumn,groupedaddress]{revtex4-1}

\usepackage[utf8]{inputenc}
\usepackage{graphicx}
\usepackage{flafter}

\usepackage{amsmath,amssymb,amstext}
\usepackage{braket}

\usepackage{float}
\usepackage{color}
\usepackage{subfigure}
\usepackage{blindtext}
\usepackage{marginnote}

\newcommand{\bk}[1]{\left(#1\right)}					
\newcommand{\Bk}[1]{\left[#1\right]}					
\newcommand{\bvec}[1]{\boldsymbol{#1}}					
\newcommand{\euler}[1]{\text{e}^{#1}}					
	
\begin{document}


\title{Floquet scattering of light and sound  in Dirac optomechanics}


\author{C. Wurl}
\email{wurl@physik.uni-greifswald.de}
\author{H. Fehske}
\email{fehske@physik.uni-greifswald.de}
\affiliation{Institut f{\"u}r Physik,
Universit{\"a}t Greifswald, 17487 Greifswald, Germany }


\date{\today}

\begin{abstract}
The inelastic scattering and conversion process between photons and phonons by laser-driven quantum dots  is analyzed for a honeycomb array of  optomechanical cells. Using  Floquet theory for an effective two-level system, we solve the  related time-dependent scattering problem, beyond the standard rotating-wave approximation approach, for a plane  Dirac-photon wave hitting a cylindrical oscillating barrier that couples the radiation field to the vibrational degrees of freedom. We demonstrate different scattering regimes and discuss the formation of polaritonic quasiparticles. We show that sideband-scattering becomes important when the energies of the sidebands are located in the vicinity of avoided crossings of the quasienergy bands. The interference of Floquet states belonging to different sidebands causes a mixing of long-wavelength (quantum) and short-wavelength (quasiclassical) behavior, making it possible to use the oscillating quantum dot as a kind of transistor for light and sound. We  comment under which conditions  the  setup  can be utilized to observe \textit{zitterbewegung}.
\end{abstract}


\maketitle

\section{Introduction}
Optomechanical systems realizing the interaction between light and matter on the micro- and macroscale~\cite{AKM14}, enjoy continued interest since they allow for the study of fundamental questions  concerning, e.g.,  the cooling of nanomechanical oscillators into the quantum groundstate~\cite{Chea11,Teea11,FGN16}, nonlinear phenomena on the route from classical~\cite{MHG06,WAF16} to quantum behavior~\cite{BAF15,QCHM12,CABF16}, and even entanglement~\cite{Viea07,GKPBLS14} and (quantum) information processing~\cite{WC12,HS12,PHTSL13,WB17}. Regarding the latter one, optomechanical crystals or arrays~\cite{ECCVP09,SP10,SMWP10,SA14} have  gained particular attention as they  accommodate (strongly) coupled collective modes~\cite{HL11,XGD12,LM13}, and therefore can be utilized for the transport, storage, and transduction of photons and phonons~\cite{CS11,SP11,SLM12,CC14,FMLP16}. 

A promising building block for  hybrid photon-phonon signal processing architectures is provided by planar optomechanical metamaterials. Their optically tunable, polaritonlike band structure enables versatile and easy to implement applications of artificial optomechanical gauge fields~\cite{SK15,WM16,ANG18} and topological phases  of light and sound~\cite{PBSM15}. In this context, the emergence of Dirac physics was demonstrated for low-energy photons and phonons in "optomechanical graphene", that is, a honeycomb array of optomechanical cells~\cite{SPM15}.  In these systems ultrarelativistic transport phenomena such as Klein tunneling  appear, because of the chiral nature of the quasiparticles and their Dirac-like band structure, just as for Dirac electrons in graphene. Moreover, the radiation pressure that induces the coupling between photons and phonons inside the optomechanical  barrier can be easily tuned   by the laser power and may cause the formation of (photon-phonon) polariton states mixing photonic and phononic contribution. 
Circular barriers are of special interest because they are easier to implement experimentally than infinite planar barriers and show a richer scattering behavior due to their finite size. In particular such optomechanical "quantum dots" may cause the spatial and temporal trapping, Veselago lensing, a depletion of Klein tunneling, and  angle-dependent interconversion of photons and phonons~\cite{WF17}.

Since transport of Dirac quasiparticles is extremely energy-sensitive, external time-dependent fields may produce interesting effects. This has been demonstrated for the photon-assisted transport in graphene-based nanostructures~\cite{PA04}, where planar and circular electromagnetic potentials,  oscillating with frequency $\Omega$, give rise to inelastic scattering processes by exchanging energy quanta  $n\hbar \Omega$ with the oscillating field. Thereby, the excitation into and interference between sideband states may cause  the suppression of (Klein-) tunneling, Floquet-Fano resonances, as well as highly  anisotropic angle-resolved transmission and emission of the quasiparticles ~\cite{FE07,ZS08,LW12,SB12,BS13,SHF15_2}. Also the relevance to  \textit{zitterbewegung} (ZB)  has been addressed within the Tien-Gordon setup~\cite{TB07}. 

As stressed already, inside the optomechanical barrier polaritonic quasiparticles will form. They can be treated effectively as two-level systems. Then, modulating the coupling strength in a time-periodic way, the system mimics  a two-level system driven by a linear polarized laser field.  Within Floquet theory, it was shown that such  systems exhibit strongly enhanced transmission probabilities between the two levels whenever avoided crossings occur in the quasienergy bands~\cite{S65,SS09,BC15}. This immediately raises the question how Floquet-driven barriers affect the two-level scattering process in optomechanical metamaterials. For  planar oscillating barriers we found that the finite transmission probabilities for the sidebands might suppress or revive the light-sound interconversion when the energy of the incident photon is close to multiples of the oscillation frequency~\cite{WF18}. 

\begin{figure}
\includegraphics[width=0.48 \textwidth]{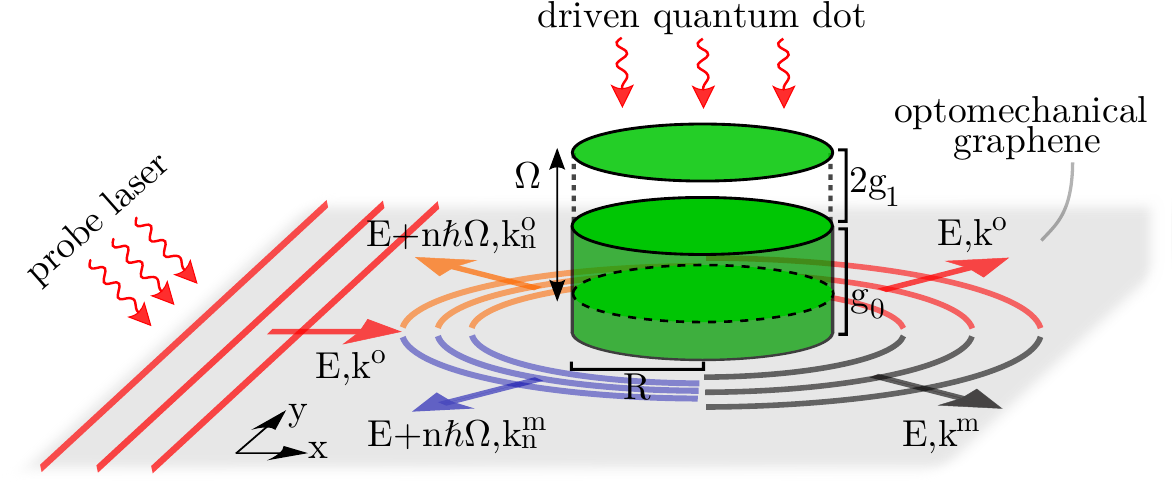}
\caption{(Color online) Sketch of the scattering setup. Injected by a probe laser, an incident optical wave with energy $E>0$ and wave vector $\bvec{k}^{o}=k^{o}\bvec{e}_{x}$ hits a laser-driven quantum dot of radius $R$. Inside the dot, the photon-phonon coupling is  $g=g_{0}-2g_{1}\cos \bk{\Omega t}$. As a result, reflected optical and mechanical waves appear  with wave vectors $\bvec{k}^{o/m}$ ($\bvec{k}_{n=0} \equiv \bvec{k}$),  in the central band with energy $E$, and in the sidebands with energies  $E_{n}=E+n\hbar \Omega$, where $n=\pm 1,\pm 2, ...$. The reflected waves are directed away from the dot and carry an angular momentum. Since the dot allows for the conversion between light and sound, mechanical waves appear outside the dot even though  the coupling vanishes here. Note that the figure is not true to scale and, since $v_{o} > v_{m}$, the photon-phonon wavevectors   $k^{o/m}_{n}=|E_{n}|/\hbar v_{o/m}$ are not equal in magnitude. \label{fig1}}
\end{figure}

Motivated by these findings, in the present paper  we study the inelastic scattering and conversion process between photons and phonons triggered by periodically oscillating quantum dots, imprinted optically in optomechanical graphene. Figure~\ref{fig1} illustrates the setup under consideration. The paper is organized as follows. Section~\ref{theo} 
presents our model and outlines the theoretical approach,  based on Floquet theory for an effective two-level system. The  solution of the related time-dependent scattering problem is explicitly given. A  more detailed presentation of  the (numerical)  implementation of our Floquet approach can be found in the appendix~\ref{app}. In Sec.~\ref{results}, after briefly recapitulating previous findings for the static quantum dot,  we discuss the numerical results obtained for the oscillating quantum dot in the whole range of system parameters. The relevance for observing ZB is also considered. Our main conclusions  can be found in Sec.~\ref{conc}.

\section{Theoretical approach}\label{theo}
\subsection{Model}
In optomechanical graphene, driven by a laser with frequency $\omega_{las}$, co-localized cavity photon (eigenfrequency $\omega_{o}$) and phonon (eigenfrequency $\omega_{m}$) modes interact via radiation pressure.  For sufficiently low energies and barrier potentials that are smooth on the scale of the lattice constant but sharp on the scale of the de Broglie wavelength (i.e., the size of the dot is much bigger than the lattice spacing in the optomechanical array), the continuum approximation applies~\cite{RM18}. Then the system can be described by the optomechanical Dirac-Weyl Hamiltonian~\cite{SPM15},
\begin{equation}\label{H}
H=\bk{\overline{v}+\frac{1}{2}\delta v \, \tau_{z}}\bvec{\sigma}\cdot\bvec{k}-g\bk{\bvec{r},t}\tau_{x}.
\end{equation}
In Eq.~\eqref{H},  the model Hamiltonian is written in units of $\hbar$, after rescaling $H\rightarrow H -\hbar\omega_{m}$. Here, $\overline{v}=(v_{o}+v_{m})/2$, $\delta v=v_{o}-v_{m}$, with $v_{o/m}$ as the Fermi velocity of the optical or mechanical mode, $\bvec{\tau}$ and $\bvec{\sigma}$ are Pauli spin matrices, $\bvec{k}$ ($\bvec{r}$) gives the wave vector (position vector) of the Dirac wave, and $g\bk{\bvec{r},t}$ parametrizes the time-dependent photon-phonon  coupling strength. On the other hand, when the laser continuously drives a certain region of the honeycomb lattice, a quantum barrier with time-independent coupling strength $g_{0}$ is created. 

We note that the above single-valley Hamiltonian is obtained  after linearizing the dynamics around the steady-state solution and taking advantage of  the rotating-wave approximation (RWA) in the red detuned moderate-driving regime, $\Delta=\omega_{las}-\omega_{o}=-\omega_{m}$~\cite{SPM15}.  
To account for   inelastic scattering, we assume the laser amplitude to be modulated with a frequency much smaller than the frequencies of both the laser and mechanical modes,  $\Omega \ll \omega_{las},\omega_{m}$ (otherwise the RWA is not granted). Furthermore, $\Omega$ should be much smaller than the  mechanical hopping in the array, i.e., $\Omega < 2v_{m}/3a$  with $a$ as the lattice constant (otherwise the continuum approximation is not granted)~\cite{SPM15}.  Then, using polar coordinates, the photon-phonon coupling in the quantum dot region with radius $R$ takes the form,
\begin{equation} \label{potential}
g\bk{r,t}=\Bk{g_{0}-2g_{1}\cos \bk{\Omega t}}\Theta\bk{R-r},
\end{equation}
where $g_{0}>0$ and $g_{1}<0$, and $g_{0,1}$ are assumed to be constant. Furthermore, in order to ensure a laser amplitude greater than zero,  $2|g_{1}|\leq g_{0}$.  In what follows, for the sake of simplicity, the potential barrier~\eqref{potential}  is assumed to be infinitely sharp. Numerical studies have shown that a more realistic steep but rounded barrier will influence the results little (due to the small Umklapp scattering)~\cite{SPM15}.

At this point we should mention that the Hamiltonian~\eqref{H}, derived for the linear regime within the RWA, takes into account dissipation effects in an effective way~\cite{AKM14,SPM15}. Accordingly, the quasiparticles  described by the model~\eqref{H} propagate as undamped optical and mechanical excitations on the honeycomb lattice. As shown in Ref.~\cite{SPM15} the main effect of dissipation would be the decay of the field amplitudes. For the same reason, the barrier is described by the optomechanical coupling strength $g$ (being proportional to the laser amplitude) and not by the single-photon coupling rate. 

Inside the quantum dot, where the photon-phonon coupling is finite, the polariton quasiparticle states are superpositions of optical and mechanical eigenstates of $\tau_{z}$. Given the time-periodic coupling~\eqref{potential}, the polariton states can be treated as periodically driven two-level systems. A similar approach is widely used in quantum optics (Rabi model), e.g., in order to model atoms or superconducting qubits driven by a semiclassical, linearly polarized laser field (see Ref.~\cite{GH98} and references cited therein).  There it is convenient to obtain the time-dependent solutions within the RWA, which is justified for laser frequencies close to the transition frequency between the two energy levels of the  state. In view of solving the scattering problem, however,  the RWA cannot be applied because the wave number $k$, which enters the transition frequency between the two polariton states, $ \delta v  k /2$ in~\eqref{H}, changes  as a result of  inelastic scattering processes. Therefore we make use of  the Floquet formalism to find the time-dependent solutions of our scattering problem. The Floquet formalism is described, e.g., in Refs.~\cite{GH98,ST04,BC15}; for its application to two-level systems see Refs.~\cite{S65,SS09,DS16}.

\subsection{Formulation of the Floquet scattering problem}
Treating the inelastic scattering problem we look for  solutions $\ket{\psi\bk{t}}$ of the time-dependent Dirac equation $i\bk{\partial/\partial t}\ket{\psi\bk{t}}=H\ket{\psi\bk{t}}$. Since the Hamiltonian is time-periodic,  according to Floquet's theorem~\cite{Floquet1883}, we write the time-dependent solution as $\ket{\psi \bk{t}}=\euler{-i\varepsilon t}\ket{\varepsilon\bk{t}}$ with quasienergy $\varepsilon$ and the time-periodic Floquet state $\ket{\varepsilon \bk{t}}=\ket{\varepsilon \bk{t+T}}$, where $T=2 \pi /\Omega$. For constructing the latter we use the eigensolutions in the absence of the oscillating barrier~\cite{WF17,SPM15}, which are given as $\ket{\tau}\ket{\sigma,\bvec{k}}$. Here, $\ket{\sigma,\bvec{k}}$ is the eigenvector of the single-particle Dirac-Weyl Hamiltonian $H=\bvec{\sigma k}$ with eigenvalue $\sigma |\bvec{k}|$ and sublattice pseudospin $\sigma=\pm 1$ (in this notation $\sigma$ acts as a band index). The  polariton state is formed according to $\ket{\tau}=\mathcal{N}^{\tau}(g_{0}\ket{o}+\gamma^{\tau}\ket{m})$, where  $\tau=\pm 1$ denotes the polariton pseudospin, and   $\ket{o/m}$ are the  bare optical and mechanical eigenstates of $\tau_{z}$ (the factors $\mathcal{N}^{\tau}$ and $\gamma^{\tau}$ are given in the appendix). Expanding the Floquet state in a Fourier series,
\begin{equation}\label{FS}
\ket{\varepsilon\bk{t}}=\sum \limits_{p}\sum\limits_{\tau=\pm}c_{p}^{\tau}\ket{\tau}\ket{\sigma,\bvec{k}}\euler{ip\Omega t}, \quad p \in \mathbb{Z}\,,
\end{equation} 
the two polariton states with $\tau=\pm 1$ have to be superimposed because of the optomechanical coupling  in $\tau$-space. Inserting the ansatz~\eqref{FS} into the time-dependent Dirac equation yields the Floquet eigenvalue equation (FEE) $\mathcal{F}\bvec{c}=\varepsilon \bvec{c}$, where $\bvec{c}$ is the vector containing the Fourier coefficients $c_{p}^{\tau}$, and $\mathcal{F}$ is  the Floquet matrix having eigenvalues $\varepsilon$.  The Floquet matrix and the FEE in component form are given in the appendix;  see Eq.~\eqref{app:FEE} and Eq.~\eqref{FEE}, respectively. In  general, an analytical solution of the FEE    does not exist~\cite{GH98}. This is in contrast to the scattering of graphene-electrons by time-periodic gate-defined potential barriers, for which the diagonal potential in sublattice space allows one to integrate the Dirac equation~\cite{PA04,TB07,ZS08,SHF15_2}. We therefore determine the solutions of the FEE numerically; see appendix.

Let us take another look at the Floquet-scattering setup depicted in Fig.~\ref{fig1}.  Since the oscillating quantum dot gives (takes) energy to (away from) photons and phonons in the form of multiple integers of the oscillation frequency, $E_{n}=E+n \Omega$ ($n \in \mathbb{Z}$), the scattering is inelastic.   This implies that  the wave functions have to be expressed as superpositions of states with energies $E_{n}$. This is certainly unproblematic outside the dot, where the coupling is zero and we can use the unperturbed eigensolutions. The transmitted wave inside the dot, however, is composed of Floquet states according to Eq.~\eqref{FS}. On that account the wave numbers $q_{n}^{(\pm)}$ and the Fourier coefficients $c^{\tau,\bk{\pm}}_{p,n}$ at each energy $E_{n}=\varepsilon^{\bk{\pm}}$ have to be determined by numerical diagonalization of the Floquet matrix $\mathcal{F}$. Note that the index $\bk{\pm}$ appears because the quasienergies are two fold degenerate  owing to  the polariton pseudospin $ \tau$.

\subsection{Solution of the Floquet scattering problem}
For this purpose, we expand the plane wave state of the incoming photon in polar coordinates,
\begin{eqnarray}\label{inc}
\ket{\psi^{in}}&=&\frac{1}{\sqrt{2}}\binom{1}{1}\euler{ik^{o}x}\ket{o}\euler{-iEt}  \nonumber \\ 
&=& \sum \limits_{n,l}\delta_{n0} \, \phi^{\bk{1}}_{n,l}\bk{k^{o}_{n}r}\ket{o}\euler{-iE_{n}t},
\end{eqnarray}
where $l \in \mathbb{Z}$ is the quantum number referring to the angular momentum. The reflected (scattered) wave consists of optical and mechanical modes, $\ket{\psi^{r}}=\ket{\psi^{r; o}}+\ket{\psi^{r; m}}$ (cf. Fig.~\ref{fig1}), with
\begin{equation}\label{ref}
\ket{\psi^{r; o/m}}=\sum \limits_{n,l}  \sqrt{\frac{v_{o}}{v_{o/m}}}r_{n,l}^{o/m} \phi^{\bk{3}}_{n,l}\bk{k^{o/m}_{n}r}\ket{o/m}\euler{-iE_{n}t}.
\end{equation}
Here, $r_{n,l}^{o/m}$ are the optical/mechanical reflection coefficients. According to Eq.~\eqref{FS}, the transmitted wave $\ket{\psi^{t}}=\ket{\psi^{t;\bk{+}}}+\ket{\psi^{t;\bk{-}}}$ reads
\begin{eqnarray}\label{trans}
\ket{\psi^{t;\bk{\pm}}}&=&\sum \limits_{n,l} t_{n,l}^{\bk{\pm}}\phi_{n,l}^{\bk{1}}\bk{q_{n}^{\bk{\pm}}r} \nonumber \\
&\times & \sum \limits_{p}\sum\limits_{\tau=\pm}c_{p,n}^{\tau,\bk{\pm}}\ket{\tau}^{\bk{\pm}}_{n}\euler{-iE_{n-p}t}\,,
\end{eqnarray}
where $t^{\bk{\pm}}_{n,l}$ are the transmission coefficients. The Fourier coefficients and  wave numbers  used in Eq.~\eqref{trans} are extracted from the Floquet approach outlined in the appendix. For the wavefunctions~\eqref{inc}--\eqref{trans} we have used the eigenfunctions $\braket{\bvec{r}|\sigma,\bvec{k}}$ of the Dirac-Weyl Hamiltonian~\cite{HBF13,SHF15,SHF15_2},
\begin{equation}
\phi_{n,l}^{\bk{1,3}}\bk{k_{n}r}=\frac{1}{\sqrt{2}}i^{l+1}\begin{pmatrix}
-i \mathcal{Z}_{l}^{\bk{1,3}}\bk{k_{n}r}\euler{il\varphi} \\
 \sigma_{n} \mathcal{Z}_{l+1}^{\bk{1,3}}\bk{k_{n}r}\euler{i(l+1)\varphi}
\end{pmatrix},
\end{equation}
where $\mathcal{Z}^{\bk{1}}=J_{l}$ and $\mathcal{Z}^{\bk{3}}=H_{l}$ denotes the Bessel function and Hankel function, respectively. To ensure that the group velocity of the reflected wave is directed away from the quantum dot (as it should be for an outgoing wave), the sign of the energy determines which kind of Hankel function is used: $H_{l}=J_{l}+i\sigma_{n}^{out}Y_{l}$ ($Y_{l}$ is the Neumann function). Here, $\sigma_{n}^{out}=\text{sgn}\bk{E_n}$ is the 'band index' outside the quantum dot. Its presence in the Hankel function ensures that the refractive indices are negative for negative energies, meaning  that   the wave vector  is directed opposite  the propagation direction of the particle. For the transmitted wave inside the dot, $\sigma_{n}^{ins}=\pm 1$ for $E_{n}\gtrless \pm g_{0}$, and $\sigma^{ins \bk{\pm}}_{n}=\pm 1$ for $-g_{0}\leq E_{n} \leq g_{0}$. Matching the wave functions at $r=R$ yields the equations for the transmission coefficients:
\begin{subequations}\label{scatmatrix1}
\begin{align} 
\delta _{p0}W_{p,l}^{o} &=\sum \limits _{n}\sum \limits_{\tau=\pm}t_{n,l}^{\bk{\tau}}f_{n-p,n}^{\bk{\tau}}X_{n,p,l}^{o,\bk{\tau}},\\ 
0 &=\sum \limits _{n}\sum \limits_{\tau=\pm}t_{n,l}^{\bk{\tau}}h_{n-p,n}^{\bk{\tau}}X_{n,p,l}^{m,\bk{\tau}}.
\end{align}
\end{subequations}
The reflection coefficients can be obtained from
\begin{subequations}\label{scatmatrix2}
\begin{align} 
r_{p,l}^{o}= & \sum \limits_{n}\sum \limits_{\tau=\pm} t_{n,l}^{\bk{\tau}}f_{n-p,n}^{\bk{\tau}}\frac{\mathcal{Z}_{l}^{\bk{1}}(q_{n}^{\bk{\tau}}R)}{\mathcal{Z}_{l}^{\bk{3}}(k_{p}^{o}R)}-\delta_{p0} \frac{\mathcal{Z}_{l}^{\bk{1}}(k_{p}^{o}R)}{\mathcal{Z}_{l}^{\bk{3}}(k_{p}^{o}R)}, \\ 
r_{p,l}^{m}= & \sum \limits_{n}\sum \limits_{\tau=\pm} t_{n,l}^{\bk{\tau}}h_{n-p,n}^{\bk{\tau}}\frac{\mathcal{Z}_{l}^{\bk{1}}(q_{n}^{\bk{\tau}}R)}{\mathcal{Z}_{l}^{\bk{3}}(k_{p}^{m}R)}.
\end{align}
\end{subequations}
Here, we have used the abbreviations
\begin{subequations}
\begin{align}
W_{p,l}^{o}=& \phantom{-,}\, \mathcal{Z}^{\bk{1}}_{l}(k_{p}^{o}R)\mathcal{Z}^{\bk{3}}_{l+1}(k_{p}^{o}R) \nonumber \\ 
& -\mathcal{Z}^{\bk{1}}_{l+1}(k_{p}^{o}R)\mathcal{Z}^{\bk{3}}_{l}(k_{p}^{o}R), \\
X_{n,p,l}^{o/m,\bk{\tau}}=& \phantom{-,}\, \sigma_{p}^{out}\mathcal{Z}^{\bk{1}}_{l}(q_{n}^{\bk{\tau}}R)\mathcal{Z}^{\bk{3}}_{l+1}(k_{p}^{o/m}R) \nonumber \\ 
& -\sigma^{ins\bk{\tau}}_{n}\mathcal{Z}^{\bk{1}}_{l+1}(q_{n}^{\bk{\tau}}R)\mathcal{Z}^{\bk{3}}_{l}(k_{p}^{o/m}R),
\end{align}
\end{subequations}
and 
\begin{subequations}
\begin{align}
f_{n-p,n}^{\bk{\tau}} &=  \sum\limits_{\tau '} c_{n-p,n}^{\tau',\bk{\tau}}\mathcal{N}^{\tau ',\bk{\tau}}_{n}g_{0}, \\
h_{n-p,n}^{\bk{\tau}} &= \sum\limits_{\tau '} c_{n-p,n}^{\tau',\bk{\tau}}\mathcal{N}^{\tau ',\bk{\tau}}_{n}\gamma_{n}^{\tau',\bk{\tau}}.
\end{align}
\end{subequations}
When solving the infinite-dimensional coupled linear system~\eqref{scatmatrix1} numerically, we raise the dimension of the coefficient (scattering) matrix until convergence is reached. This is most challenging for large $g_{1}$ or small $\Omega$, since  the dimension of the scattering matrix is mainly determined by the ratio $|g_{1}|/\Omega$ (cf. appendix).

The inelastic scattering and conversion process between photons and phonons is characterized by the scattering efficiency $Q^{o/m}(r,t)$, that is, the scattering cross section divided by the geometric cross section. 
It consists of a time-averaged part
\begin{equation}\label{Qa}
\overline{Q}^{o/m}=\sum \limits_{n}\sum \limits_{l=0}^{\infty}\overline{Q}^{o/m}_{n,l}=\sum \limits_{n}\sum \limits_{l=0}^{\infty}\frac{4}{k_{n}^{o/m}R}\left|r_{n,l}^{o/m}\right|^2,
\end{equation}
and a time-dependent part (to simplify the notation, we omit the index $out$ in $\sigma^{out}_{n}$)
\begin{align}\label{Qt}
&\tilde{Q}^{o/m}\bk{r,t}= \sum \limits_{n<n'}\sum \limits_{l=0}^{\infty} (-1)^{l}\frac{4}{\sqrt{k_{n}^{o/m}k_{n'}^{o/m}}R} \nonumber \\
& \times 2\mathfrak{R}\left\{(r_{n',l}^{o/m})^{*}r_{n,l}^{o/m} i^{\frac{1}{2}(\sigma_{n'}-\sigma_{n})}\, \euler{i\bk{n-n'}\Omega \vartheta^{o/m}}\right\}\,.
\end{align}
Here,  $\vartheta^{o/m}=r/v_{o/m}-t$ denotes the time-retarded phase factor. In Eqs.~\eqref{Qa}, \eqref{Qt},  and hereafter, $l \geq 0$.  The quantities $\overline{Q}_{n,l}^{o/m}$ in Eq.~\eqref{Qa} represent the scattering contributions of the partial wave $l$ and the sideband $n$.  In the far field, the scattering efficiency is obtained from the radial component of the current density of the reflected wave, $(1/2R)\int j_{r}^{r;o/m}\bk{r,t} r \text{d}\varphi$~\cite{HBF13,SHF15,SHF15_2,WF17},
\begin{align}\label{farfield}
&j_{r}^{r;o/m}\bk{r,t}=\sum\limits_{n,n'}\sum \limits_{l,l'}\frac{4v_{o}}{\pi \sqrt{k_{n}^{o/m}k_{n'}^{o/m}}r}(r_{n',l'}^{o/m})^{*}r_{n,l}^{o/m} \nonumber \\
& \times   i^{l-l'}i^{\frac{1}{2}(\sigma_{n'}-\sigma_{n})}i^{(l+l')\text{sgn}(\sigma_{n'}-\sigma_{n})+(l'-l)\text{sgn}(\sigma_{n'}+\sigma_{n})} \nonumber \\
& \times  \big\{\cos \Bk{\bk{l+l'+1}\varphi} + \cos \Bk{\bk{l-l'}\varphi}\big\} \euler{i\bk{n-n'}\Omega \vartheta^{o/m}},
\end{align}
which characterizes the angular scattering. In the near-field,  the scattering is further specified  by the probability density $\rho=\braket{\psi|\psi}$, with $\ket{\psi}=\ket{\psi^{in}}+\ket{\psi^{r}}$ outside and $\ket{\psi}=\ket{\psi^{t}}$ inside the quantum dot. Note that in the far-field, the optical/mechanical part of the probability density of the reflected wave $\braket{\psi^{r}|\psi^{r}}$  becomes equal to the current density~\eqref{farfield} except for a constant factor $v_{o/m}$.  Furthermore,  defining the scattering efficiency by the cross section,  only  the incident current of the photon was used, since no phonon incident currents exist (cf. Fig.~\ref{fig1}). Therefore, the scattering efficiency of the phonon $Q^{m}$ can be understood as an interconversion rate between photons and phonons, which we can define as  $Q^{m}/Q^{o}$. 
\section{Numerical results}\label{results}
Since the scattering problem worked out in the preceding section is invariant under the transformation $[E, g_{0,1} , \Omega, R^{-1}] \rightarrow \gamma[E, g_{0,1} , \Omega, R^{-1}]$ with $\gamma \in \mathbb{R}$, we rescale the equations of motion such that  $\Omega=1$~\cite{WF18}. We set $v_{o}=10v_{m}$ and furthermore employ units such that $v_{o}=\hbar=1$~\cite{SPM15, WF17,WF18}. Then, the rescaled variables are dimensionless and  related to the unscaled variables (marked by $\hat{\phantom{a}}$) according to $E=\hat{E}/(\hbar \Omega)$, $g_{0,1}=\hat{g}_{0,1}/\Omega$, $R=\hat{R}\Omega/v_{o}$, $k=\hat{k}v_{o}/\Omega$.  The phase factor is measured in units of $\Omega$, $\vartheta^{o/m}=\hat{\vartheta}^{o/m}\Omega$. According to the experimental parameters given in Ref.~\cite{SP10} the effects discussed in this paper should be observable  for oscillation frequencies $\Omega \sim 0.5\text{MHz} \ll \omega_{las}$, where we have assumed a laser-enhanced optomechanical coupling strength  $\hat{g}_{0}\sim 0.1 \text{MHz}$ with 2$|\hat{g}_{1}|\lesssim \hat{g}_{0}$. Then, without violating the continuum approximation, the energies of the photon and the phonon are in the order of $\hbar \omega_{m}$ (microwaves) with excitation energies   $ n \Omega \sim \text{MHz} \ll  \omega_{m}$ for the sidebands.  The typical size of the quantum dot radius is $100a$  with lattice constant $a\sim 50\mu \text{m}$. Using these parameters  the photon tunneling rate $J$ between two sites~\cite{SPM15}  has to be made small by design: $J=2v_{o}/3a\sim 10^{-2}\omega_{m}$.

\subsection{Static quantum dot}\label{static}

\begin{figure}
\includegraphics[width=0.48\textwidth]{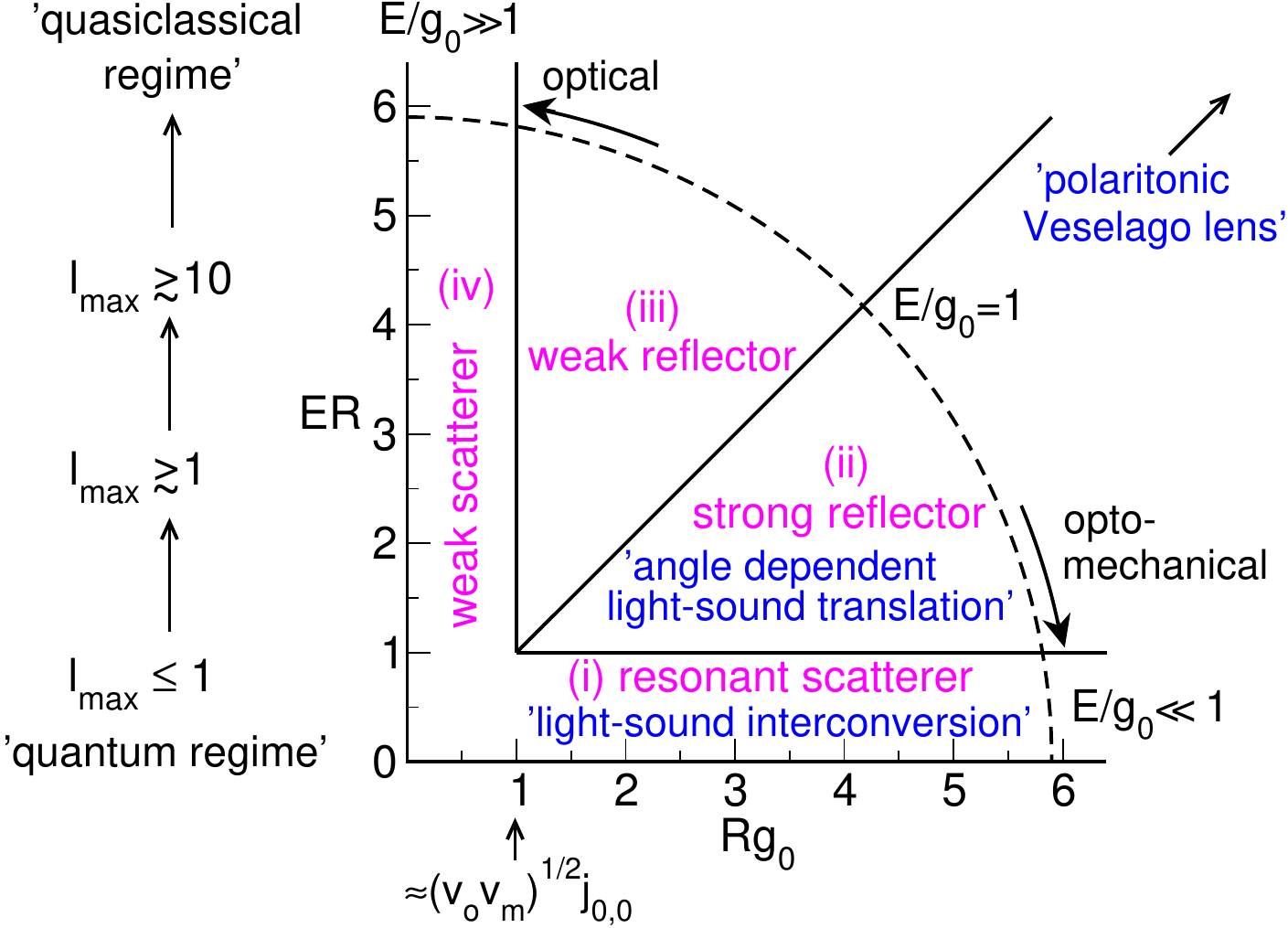}
\caption{(Color online) Different scattering regimes for the static quantum dot  in dependence on the strength parameter $Rg_{0}$ and the size parameter $ER$. The latter determines the maximum angular momentum $l_{max}$ being possible in the scattering. The energy-coupling ratio $E/g_{0}$ switches between the optomechanical ($E/g_{0} \ll 1$) and the pure optical regime ($E/g_{0} \gg 1$), where the optomechanical (optical) regime is characterized by the interconversion rate $Q^{m}/Q^{o} \sim 1$ ($\ll 1$). Depending on these parameter ratios the static dot acts as a (i) resonant scatterer in the quantum regime, (ii) strong reflector, (iii) weak reflector, or (iv) weak scatterer. On the axis of abscissae the first resonance point derived from the resonance condition~\eqref{resonance} with $l=0$  is marked.}\label{fig2}
\end{figure}

The scattering problem of the static dot ($g_{1}=0$) has  been analyzed  in  previous work~\cite{WF17}. Depending  on the strength parameter $Rg_{0}$ and  the size parameter $ER$, different scattering regimes occur. They can be  characterized by the scattering efficiency;  see Fig.~\ref{fig2}. This schematic figure is taken as a starting point, helping us to classify the different parameter regimes and expected physical phenomena in the theoretical discussion below.

Comparing the scattering regimes of our optomechanical quantum  dot (Fig.~\ref{fig2}) with those of electrons in graphene scattered by gate-defined quantum dots (cf. Fig. 3 in Ref.~\cite{WF14}), strong similarities could be identified, which perhaps is not surprising in view of the close relation between both Hamiltonians. The most crucial difference is the nondiagonal optomechanical coupling, which allows the quantum dot to translate  light into sound. The interconversion rate $Q^{m}/Q^{o}$ is determined by the energy-coupling ratio $E/g_{0}$ (see Fig. 3 in~\cite{WF17}) and discriminates between the optomechanical and purely optical regimes (dashed line in Fig.~\ref{fig2}).  For $E/g_{0} \ll 1$, i.e., in the resonant scattering  (quantum) regime,  the size parameter is small for not too large radii ($ER \ll  1$), so the excitation of the first partial waves leads to sharp resonances in the scattering efficiency of the photon, and of the phonon accordingly. The resonance condition is
\begin{equation}\label{resonance}
Rg_{0}=\sqrt{v_{o}v_{m}}j_{l,i},
\end{equation}
where $j_{l,i}$ denotes the $i$'th zero of the Bessel function $J_{l}$ with $i=0,1,2,...$ (the onset of the resonant scattering regime is marked by an arrow in Fig.~\ref{fig2}).  Resonances are featured by quasi-bound states in the quantum dot and preferred scattering directions in the far-field (cf., Fig. 4 in~\cite{WF17}). Increasing $E/g_{0}$ the phonon is hardly scattered and the scattering becomes weaker.  In the limit $E/g_{0} \gg 1$, the scattering becomes purely photonic because $v_{o}\ll v_{m}$. At such high photon energies  the  scattering of the phonons disappears since the corresponding refractive index is almost one.  At the same time more and more partial waves will be excited, which leads to a richer angular distribution of the radiation characteristics and the possibility of Fano resonances (cf., Figs. 5 and 6 in~\cite{WF17}). At very large size parameters, $ER \gg 1$, the wavelengths will be much smaller than the radius of the quantum dot and  the quasiclassical regime is entered. There, for $E/g_{0}<1$, the quantum dot may act as a polaritonic Veselago lens with negative refractive indices, focusing the light beam in forward direction.

\subsection{Oscillating quantum dot}\label{osci}
As already mentioned above, an oscillating quantum dot causes inelastic scattering via sideband excitations $E_{n}=E+n\Omega$ for both photons and phonons.  Hence, besides the angular momentum $l$, the sideband-energy quantum number $n$ becomes important. Accordingly the scattering regimes are no longer determined   by  $ER$ and  $E/g_{0}$, but  by  effective size parameters $E_{n}R$ and effective energy-coupling ratios $E_{n}/g_{0}$. The number of sidebands involved in the scattering is mainly determined by the ratio $|g_{1}|/\Omega$. This means, discussing the physical behavior of our setup, an additional parameter comes into play.  To avoid that the sideband-excitation energies become too large and the continuum approximation is  no longer justified possibly, in particular for the phonon with $v_{m}\ll v_{o}$,  we restrict ourselves to values of $g_{0}$ and $|g_{1}|/2$ smaller than $\Omega/2$.  \begin{figure}
\center
\includegraphics[width=0.33\textwidth]{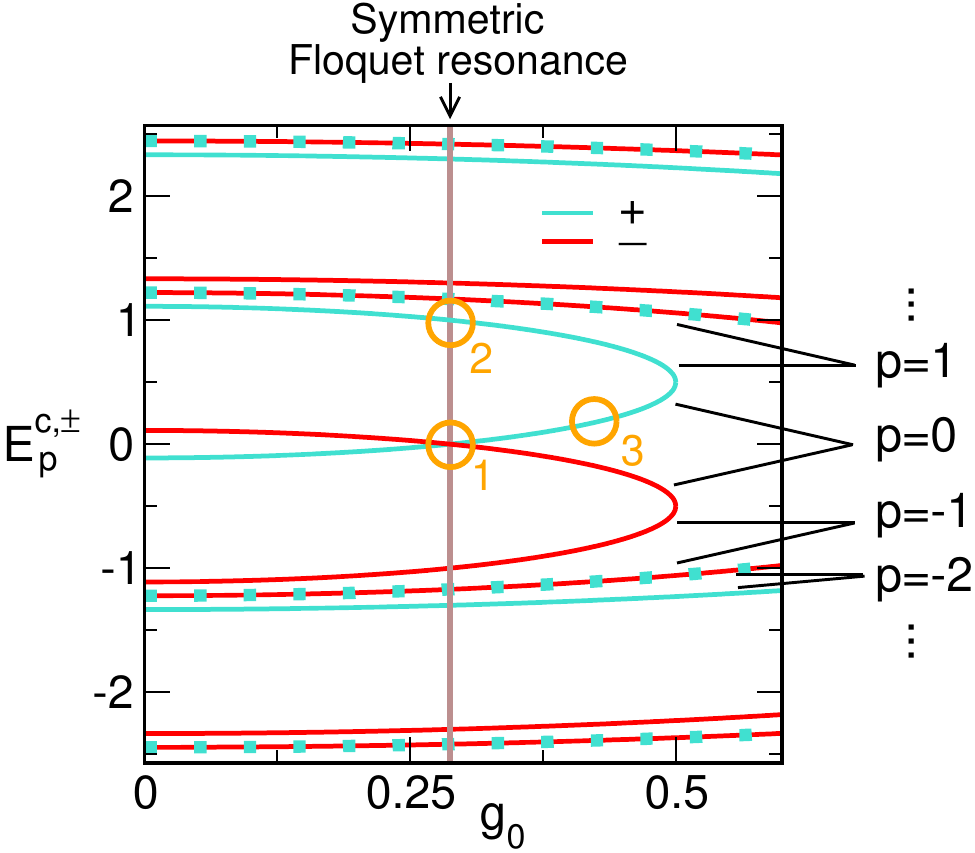}
\caption{(Color online) Crossing-energies (CE) according to Eq.~\eqref{crossings}.  
Note that in some cases CE  with different  $p$  coincide (such situation is marked by points). Symmetric Floquet-resonant scattering occurs at  $g_{0}=\Omega\sqrt{v_{o}v_{m}}/(v_{o}+v_{m}) \simeq 0.287 \Omega$. Circles designate  particular energy-coupling ratios: (1) $E/g_{0}  = 10^{-4}$ ($E\simeq 0$), (2) $E/g_{0}\simeq 3.48$ ($E=10^{-3}g_{0}+ 1\Omega\simeq  \Omega$),  and (3) $E/g_{0}\simeq 0.31$ ($E= 0.123 \Omega$ at $g_{0}= 0.394 \Omega$). They correspond  to different scattering regimes of the static dot in Fig.~\ref{fig2}: (1) $\rightarrow$ (i), (2) $\rightarrow$ (iii), (3) $\rightarrow$ (ii). \label{fig3}}
\end{figure}

Before analyzing the scattering problem in detail,  we want to make a general remark concerning our Floquet state approach. In the main,  scattering  is  determined by the refractive indices, that is to say by the different wave numbers inside and outside the scattering region. If the wave numbers inside and outside the quantum dot  are the same,   scattering disappears. The other way around, strong scattering takes place for large differences between the wave numbers belonging to the static and nonstatic cases. Clearly the deviation is  greater the larger the value of the coupling  $|g_{1}|$. Furthermore, inspecting  the quasienergies as a function of the wave number, $\varepsilon\bk{q}$, one finds the most significant deviations close to the avoided crossings (see Fig.~\ref{fig12} in the appendix). Such avoided crossings appear when two energy bands of the static case with different value of $\tau$, and  maybe shifted  by $\Omega$, cross each other. For $g_{0}\leq \Omega/2$, these crossing-energies (CE) are: 
\begin{equation}\label{crossings}
E^{c,\pm}_{p}=\pm \frac{p'}{\left|p'\right|} \frac{\bar{v}}{\delta v}\sqrt{(p'\Omega)^2-4g_{0}^2}\pm \frac{\Omega}{4}[1+(-1)^{p'+1}],
\end{equation}
with $p'=p$ for $\pm p\geq 1$  and $p'=p \mp 1 $  for $\pm p \leq  0$, where $p \in \mathbb{Z}$. Again, the polariton degree of freedom of the CE is marked  by the index $\pm$. Figure~\ref{fig3} shows the CE  depending on $g_{0}$.  Since the influence of the oscillating barrier on  the scattering is greatest for $E\sim E^{c,\pm}_{p}$, the further discussion  follows these  cases marked in Fig.~\ref{fig3}, and the subsections are numbered accordingly.

\begin{figure}
\includegraphics[width=0.48 \textwidth]{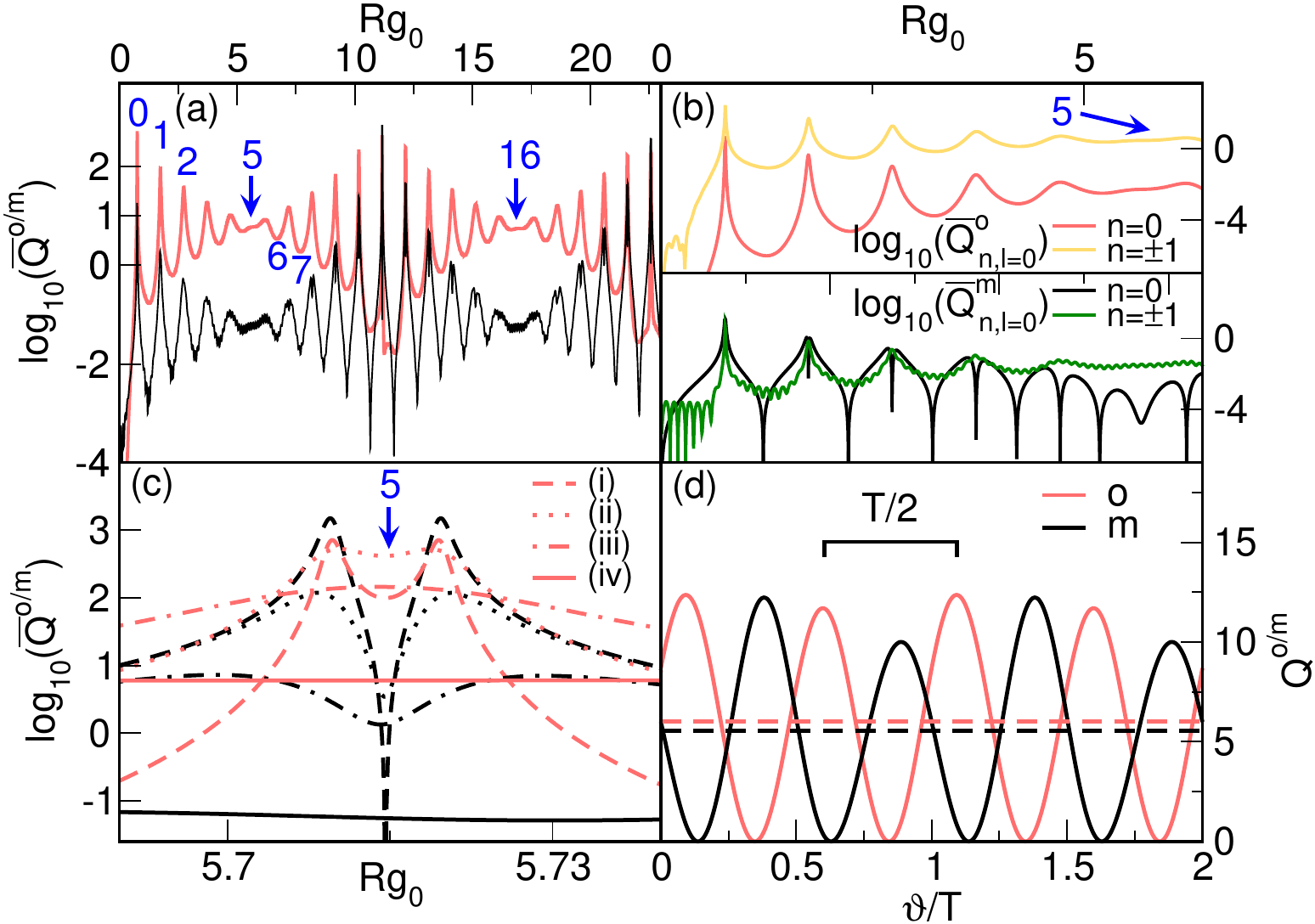}
\caption{(Color online) Scattering efficiency at weak coupling, $|g_{1}| \ll \Omega $. To realize  symmetric Floquet resonance for $E\simeq 0$ [case (1) in Fig.~\ref{fig3}] we set  $E=10^{-4}g_{0}\simeq 0$, $g_{0}=\Omega\sqrt{v_{o}v_{m}}/(v_{o}+v_{m})\simeq  0.287\Omega$, and $|g_{1}|=0.02\Omega$  [except for   (c)].  (a) Time-averaged scattering efficiency of the photon (red/gray) and the phonon (black), with resonance points $i=0,1,...$ of the static quantum dot for $l=0$ according to Eq.~\eqref{resonance} (blue numbers).  (b) Different  contributions to the scattering efficiency of  (a).  (c) Enlarged scattering efficiency close to $i=5$ for (i) $g_{1}=0$, (ii) $|g_{1}|=2\cdot 10^{-3}\Omega$, (iii) $|g_{1}|=5\cdot 10^{-3}\Omega$, and (iv) $|g_{1}|=0.02\Omega$.  (d) Time-averaged scattering efficiency (dashed) and time-evolution of the scattering efficiency at $i=5$, corresponding to case (iv) in panel (c) (here $Q_{m}$ is multiplied by a factor of 100).   \label{fig4}}
\end{figure}

\begin{figure}
\hspace{0.44cm}
\includegraphics[width=0.189 \textwidth]{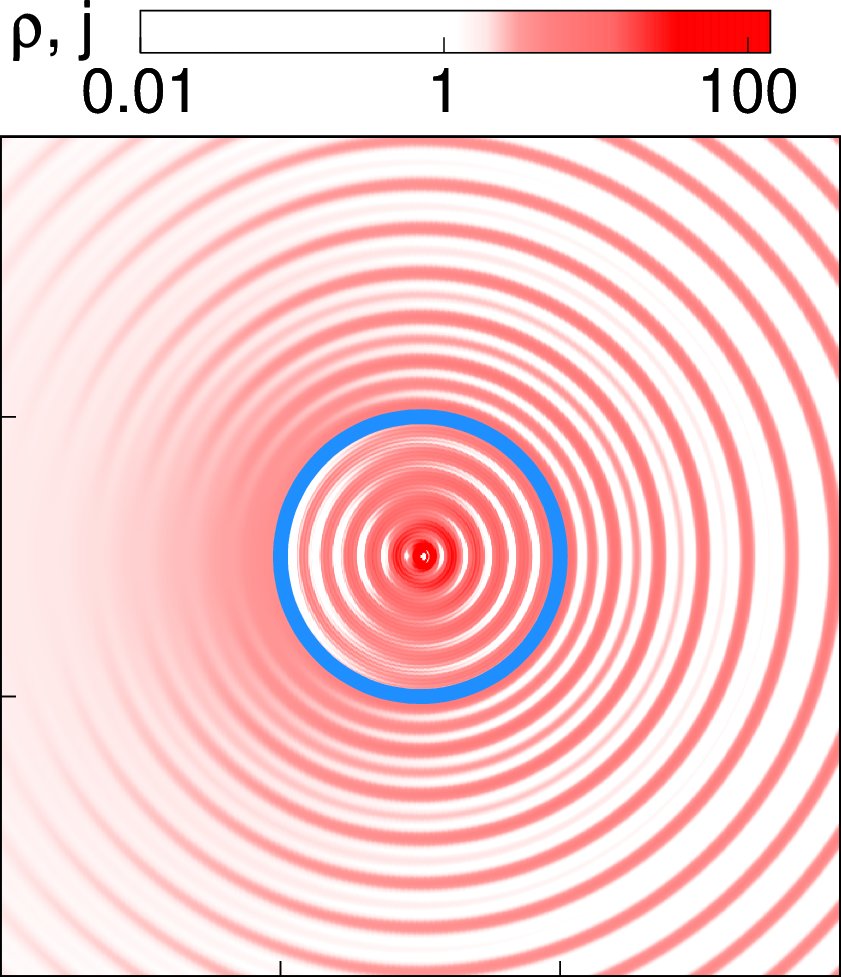}
\hspace{0.5cm}
\includegraphics[width=0.189 \textwidth]{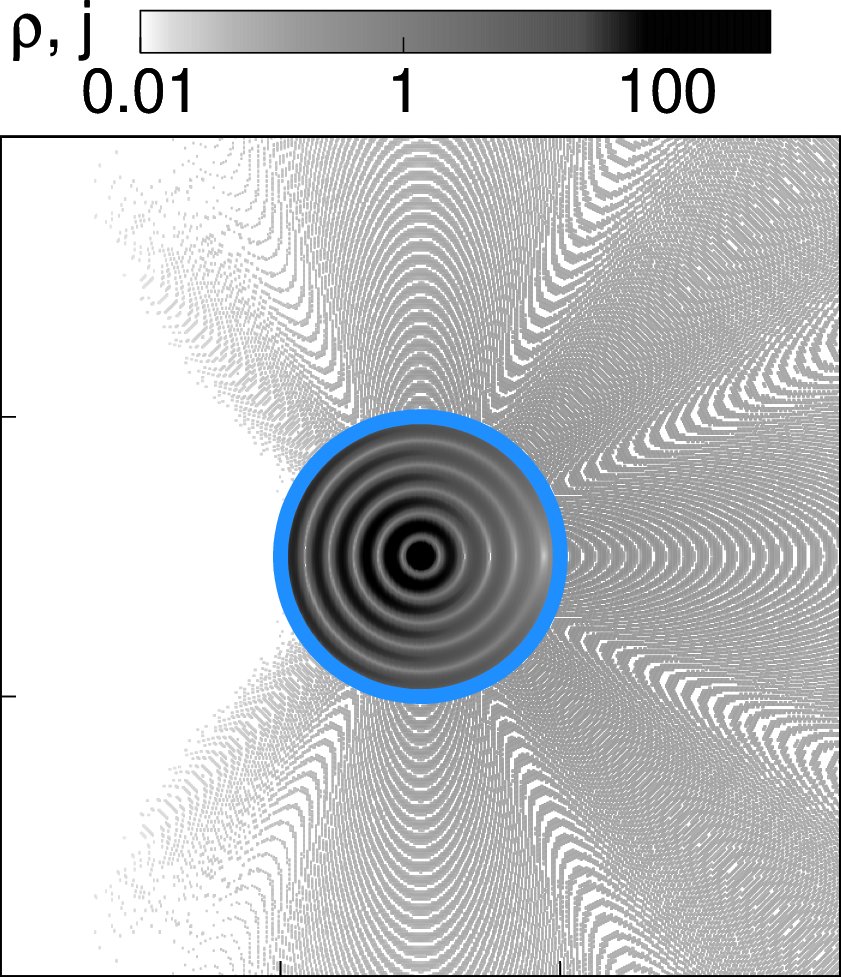} \\[0.1cm]
\includegraphics[width=0.47 \textwidth]{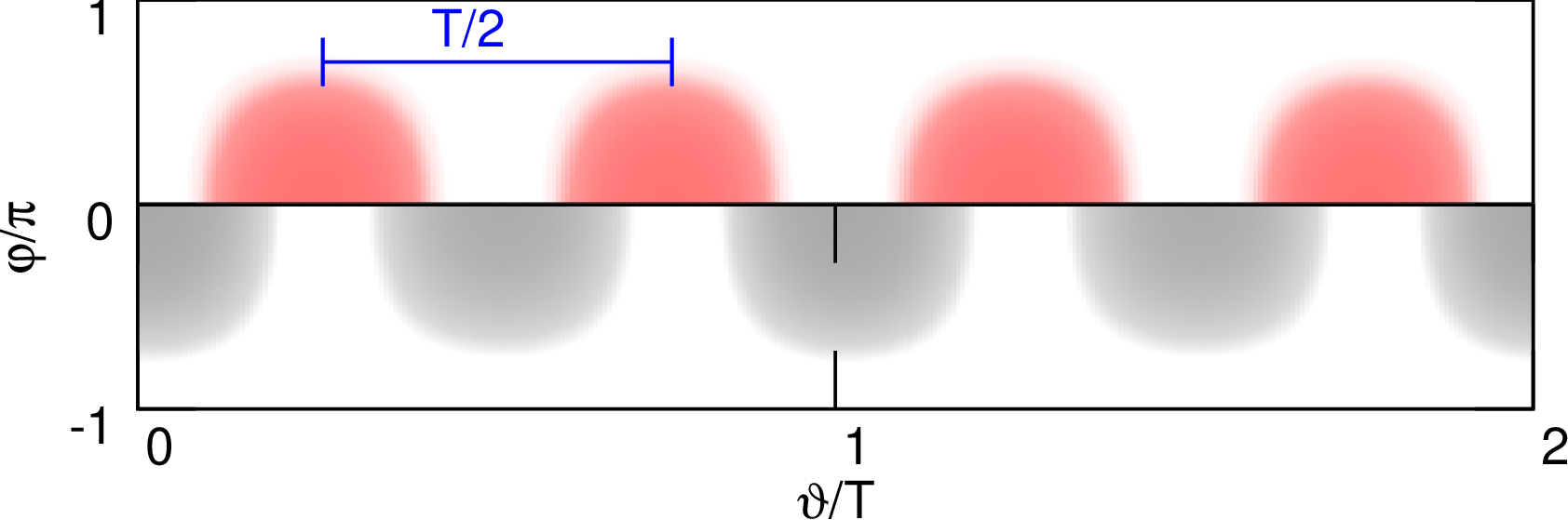}
\caption{(Color online) Time-retarded scattering characteristics for the same parameter values as used in  Fig.~\ref{fig4}(d). Shown are the optical (red, top left) and mechanical (black, top right) parts of the probability density $\rho=\braket{\psi|\psi}$ inside  and outside the quantum dot (marked by the blue circle) at $t=0$, as well as the angle-resolved time evolution of the far-field current density $j$ according to Eq.~\eqref{farfield} (bottom)  at $r=R$.   For reasons of symmetry the angle dependence of the optical (mechanical) mode is given only for $\varphi \geq 0$ ($\varphi \leq 0$). Note that the ring structure occurring in the photon probability density also exists in the phonon density, but is hard to resolve due to the small wavelength of the phonon wave ($v_{m} =v_{o}/10$) (the additional structures in the phonon density arise due to undesirable aliasing effects).   \label{fig5}}
\end{figure}

\subsubsection{Symmetric Floquet-resonant scattering close by $E\simeq 0$}
\label{symmetric}
For $g_{0}=\Omega \sqrt{v_{o}v_{m}}/(v_{o}+v_{m})$ and an incident photon energy close to the neutrality point, $E\simeq 0$ [case (1) in Fig.~\ref{fig3}], the static  dot is a resonant scatterer (quantum regime) which makes light-sound conversion possible [regime (i) in Fig.~\ref{fig2}]. Since the CE with $p=\pm 1$ are shifted by $\pm \Omega$ with respect to the $p=0$ CE and the energies $E_{n}$ are also shifted by multiples of $\Omega$ amongst themselves, we call the scattering  "Floquet-resonant". 
We find that  different CE  with $p=0$ cross at $E=0$, which entails antiparallel wave vectors of equal magnitudes inside the dot (see Fig.~\ref{fig12} in the appendix). 
In principle, the same argumentation applies to the sideband energies $E_{\pm n}$, which is why we call this  situation "symmetric". 

\paragraph{Weak photon-phonon coupling.}
Fig.~\ref{fig4} contrasts the (time-averaged) scattering efficiency of the photon and the phonon at weak couplings, i.e., in the (antiadiabatic) limit $2|g_{1}| \ll \Omega$. Obviously,  the  scattering efficiency of the static dot, with its resonances of the lowest partial wave $l=0$,  is retained to a certain extent [see Fig. 4(a)]. The resonances of the static dot can be related to minima in the scattering efficiency ($i=6,7,...$).  Most notably,  at certain points ($i=5,16,...$) the scattering is off resonant, with the result that  light-sound interconversion is strongly suppressed ($\overline{Q}^{m}/\overline{Q}^{o} \ll 1$). Although not shown here, the positions of off-resonances are moving  closer together, and towards smaller values of $Rg_{0}$, if $g_{1}$ is increased. This can be ascribed to a Fabry-P\'{e}rot interference between waves with different wave numbers inside the dot~\cite{WF18}.

Figure~\ref{fig4}(b) gives the individual contributions  to the total scattering efficiency depicted in Fig.~\ref{fig4}(a). Whereas in the static case the scattering is determined by the central band $n=0$, for finite values of $g_{1}$ the sidebands $n=\pm 1$ are involved [sidebands with $|n|>1$ (not shown) play a minor role only].  Due to the symmetry of the problem for  $E \rightarrow 0$, the  $n=\pm 1$ sideband contributions are equal in magnitude; $|r_{n=1,l}^{o/m}|\simeq|r_{n=-1,l}^ {o/m}|$.  We find that for these sidebands only the lowest partial wave with $l=0$ is excited, although the effective size parameter might suggest the opposite: $E_{n=\pm 1}R\simeq \pm \Omega R \gg 1$. We will come back to that later.  We further observe that the sidebands have large impact on the scattering, even though the coupling is weak.  This applies in particular to the off-resonance situation $i=5$, where the scattering is dominated by the sidebands  for both photons and phonons. Apparently the occurrence of off-resonances featured by weak scattering efficiency are a direct consequence of the  presence of sidebands. Since the effective energy-coupling ratio of the central band $E_{n=0}/g_{0}\simeq 0$ and the sidebands $|E_{n=\pm 1}|/g_{0} \simeq 3.48 $ lie within different scattering regimes, cf. Fig.~\ref{fig2}, their interplay may lead  to a partial transition from the resonant scattering regime to the weak reflection regime [(i)--(iii) in Fig.~\ref{fig2}], accompanied by a suppression and revival of light-sound interconversion.

\begin{figure}
\includegraphics[width=0.48\textwidth]{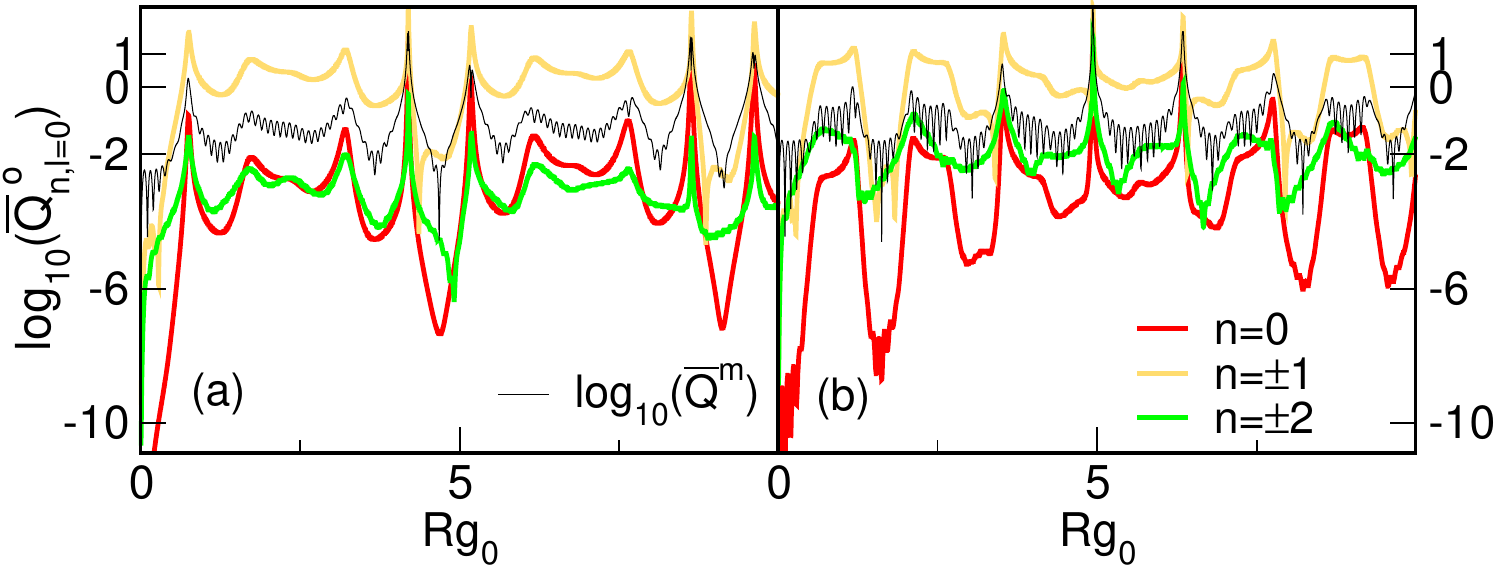}
\caption{(Color online) Scattering contributions $\overline{Q}_{n,l=0}^{o}$ [colored (thicker) lines] at moderate couplings $|g_{1}|=0.05 \Omega$  (a) and  $|g_{1}|=0.14 \Omega$ (b). Other  parameters values are the same as in Fig.~\ref{fig4} for the symmetric Floquet-resonant situation with $E\simeq 0$ [case (1) in Fig.~\ref{fig3}]. In addition, the time-averaged scattering efficiency of the phonon  is depicted [black (thin) line].   \label{fig6}}
\end{figure}

To monitor how the scattering resonance of the static dot gradually dissolves and is replaced by an off-resonance, Fig.~\ref{fig4}(c) displays the time-averaged scattering efficiency  in the vicinity of resonance point $i=5$ for different values of $g_{1}$. The resonance of the static dot [case (i)] is widely weakened for a small perturbation already [cases (ii) and (iii)], particularly for the mechanical mode. We note that the scattering resonance is characterized by two resonance peaks, occurring symmetrically about the resonance point~\cite{WF17}. At  even larger values of $g_{1}$ the resonance almost vanishes and the scattering becomes weak and purely photonic [case (iv)]. 

In Fig.~\ref{fig4}(d) the time dependent scattering efficiency is depicted at the off-resonance ($i=5$). According to Eq.~\eqref{Qt}, the sidebands ($n=\pm 1$)  interference  entails a periodic time dependence of the scattering efficiency with frequency $2\Omega$. As a result the quantum dot switches between purely photonic and phononic emission. 
In a certain sense, this time-periodic oscillation is related to ZB (but see the discussion below)~\cite{TB07}.

In Fig.~\ref{fig5} the time-retarded and periodic emission of light and sound by the oscillating quantum dot is illustrated by means of the probability density at $t=0$ (top) and the time-dependent far-field current density according to Eq.~\eqref{farfield}  at $r=R$  (bottom) for parameters of  Fig.~\ref{fig4}(d). The time periodicity of the scattering efficiency displayed in  Fig.~\ref{fig4}(d) is due to the constructive and destructive interference of the reflected wave functions for the sidebands $n=\pm 1$ and gives reason to the  ring structure with wavelength $\lambda_{o/m}=\pi v_{o/m}/\Omega$ in the probability density. For the photon density the incoming wave function covers this periodicity farther away from the dot where the wavelength is twice as large. Inside the dot the probability density  is significantly enhanced, for both photons and phonons, which can be related to the excitation of the $l=0$ mode~\cite{WF17}. Obviously, the dot captures  the incident photon and partly converts it into phonons, and emits both particle waves (periodically in time) predominantly in forward direction afterwards. In the far field, this gives rise to a time-periodic current density. The absence of backscattering at $\varphi=\pi$, related to Klein tunneling, is caused by the conservation of helicity at perpendicular incidence~\cite{SPM15,WF17} and is observed for time-dependent planar barriers as well~\cite{WF18}.

\paragraph{Moderate photon-phonon coupling.}
Figure~\ref{fig6} shows the contributions to the time-averaged scattering efficiency of the photon in this case,  where $ 2|g_{1}| \gtrsim 0.1 \Omega $.   Again only the  $l=0$ mode is noticeably excited. We find that scattering is still dominated by the sidebands with $n=\pm 1$; the contributions of the sidebands $n=\pm 2$ are rather small and are comparable with those of the central band $n=0$; see Fig.~\ref{fig6}(a). Sideband contributions with $|n|>2$ are negligible. The situation does not change much for the relatively large coupling used in Fig.~\ref{fig6}(b). The minor significance of sidebands with $|n|>1$ is obvious by looking at the CE in Fig.~\ref{fig3}: Since the sideband energies $E_{n}=E+n \Omega \simeq n \Omega$ do not match any CE for $n>1$,  these sidebands become important only at very large $g_{1}$, when the influence of the closest CE is large enough.  Figure~\ref{fig6} furthermore shows that off-resonances are still present and get closer for the higher coupling. This is  again due to interference of waves with different wave numbers inside the dot. Hence the concomitant suppression of the light-sound interconversion at the off-resonances ($\overline{Q}^{m}/ \overline{Q}^{o} \ll 1$) takes place also in the  weak resonant reflection regime.

\paragraph{Relation to zitterbewegung.}\label{zb}
In a nutshell, ZB means the rapid and tiny fluctuations of the expectation value of the particle position (velocity) about the average path due to interference of positive and negative energy states. Although the effect has never been observed for a free electron due to the largeness of its rest energy,  gapless metamaterials as  (optomechanical) graphene with its Dirac-like quasiparticles provide a promising platform to observe ZB~\cite{Katsnelson2006,TB07,MJ10,ZR11,GC14}. Let us briefly discuss the conditions under which ZB might be observable in our setup (for the moment, we set $v_{o/m}=1$).

In the absence of an oscillating barrier, $g=0$, ZB may show up in the expectation value of the velocity operator $\bvec{v}=\bvec{\sigma}$. Consider a general wave packet for the optical or the mechanical mode, respectively, given at $t=0$ as the superposition of plane wave states with positive ($\sigma=+1$) and negative energy states ($\sigma=-1$): $\ket{\psi}=(1/\sqrt{2})\sum_{\sigma}\int  a^{\sigma}(k,\varphi) \ket{\sigma,\bvec{k}}\text{d}^{2}\bvec{k}$. Here, $a^{\sigma}\bk{k,\varphi}$ is the  probability amplitude in $k$-space. Straightforward calculation in the Heisenberg picture yields $\braket{\bvec{v}}\bk{t}=\braket{\bvec{v}}_{av}+\braket{\bvec{v}}_{ZB}\bk{t}$ where  
$\braket{\bvec{v}}_{av}= \bvec{e}_{r} \frac{1}{2}\sum_{\sigma}\sigma \int \text{d}^2\bvec{k}|a^{\sigma}(\bvec{k})|^2$
is the average velocity of a free, ultrarelativistic  particle in polar coordinates and 
\begin{eqnarray}\label{velocity_ZB}
\braket{\bvec{v}}_{ZB}\bk{t}= -\bvec{e}_{\varphi} \text{Re} &\Big\{& \int \text{d}^2 \bvec{k}  \, \Bk{a^{+}\bk{k,\varphi}}^{*}a^{-}\bk{k,\varphi}\nonumber \\
&\times& \Bk{\sin \bk{2kt}-i\cos(2kt)}\Big\}
\end{eqnarray}
represents the ZB term. Equation~\eqref{velocity_ZB} clearly shows that the interference of states with positive and negative energy  is a condition for the occurrence of ZB. In addition, since the velocity operator $\bvec{\sigma}$ does not act in $k$-space $\braket{\sigma',\bvec{k}'|\bvec{\sigma}|\sigma,\bvec{k}} \sim \delta\bk{\bvec{k-k'}}$, for observing ZB, states with different helicity  have to be superimposed, i.e., the propagation directions of the states with positive and negative energy must be antiparallel.

Our results suggest that the setup considered here represents a realistic opportunity to observe ZB in optomechanics.  
Looking at the reflected wave function~\eqref{ref}, the energetic condition for ZB can be quite simply fulfilled in the case of a symmetric Floquet resonance for photon energies at the neutrality point [see Fig.~\ref{fig4}(b)]. 
Here, sideband states with positive ($E_{n=+1}\simeq +\Omega$) and negative  ($E_{n=-1}\simeq -\Omega$) energy can be symmetrically excited for both the photon and the phonon, whereby the central-band state ($E\simeq 0$)  fortunately is de-excited. The resulting  ZB frequency of  $2\Omega$ can be made small by tuning the optomechanical coupling via the laser power ($\Omega \sim g \sim 1 \text{MHz}$ by our estimates), which should be advantageous in view of an experimental implementation, just as the simple optical readout. 

We argue that the other condition can easily be fulfilled by a setup where  two optomechanical barriers (circular or planar) hit by photon waves from opposite directions, generated by the probe laser after passing a beam splitter. Then, in the space between the two barriers, where the reflected waves of either barrier interfere, ZB should be able to form (this is not the case for only one barrier, where the reflected waves have the same helicity).
A detailed analytical and numerical analysis of a suchlike extended (much more complicated) scattering problem is beyond the scope of the present work and is therefore postponed to a forthcoming study.

\subsubsection{Symmetric Floquet-resonant scattering close by $E\simeq \Omega$}
\begin{figure}
\center
\includegraphics[width=0.48\textwidth]{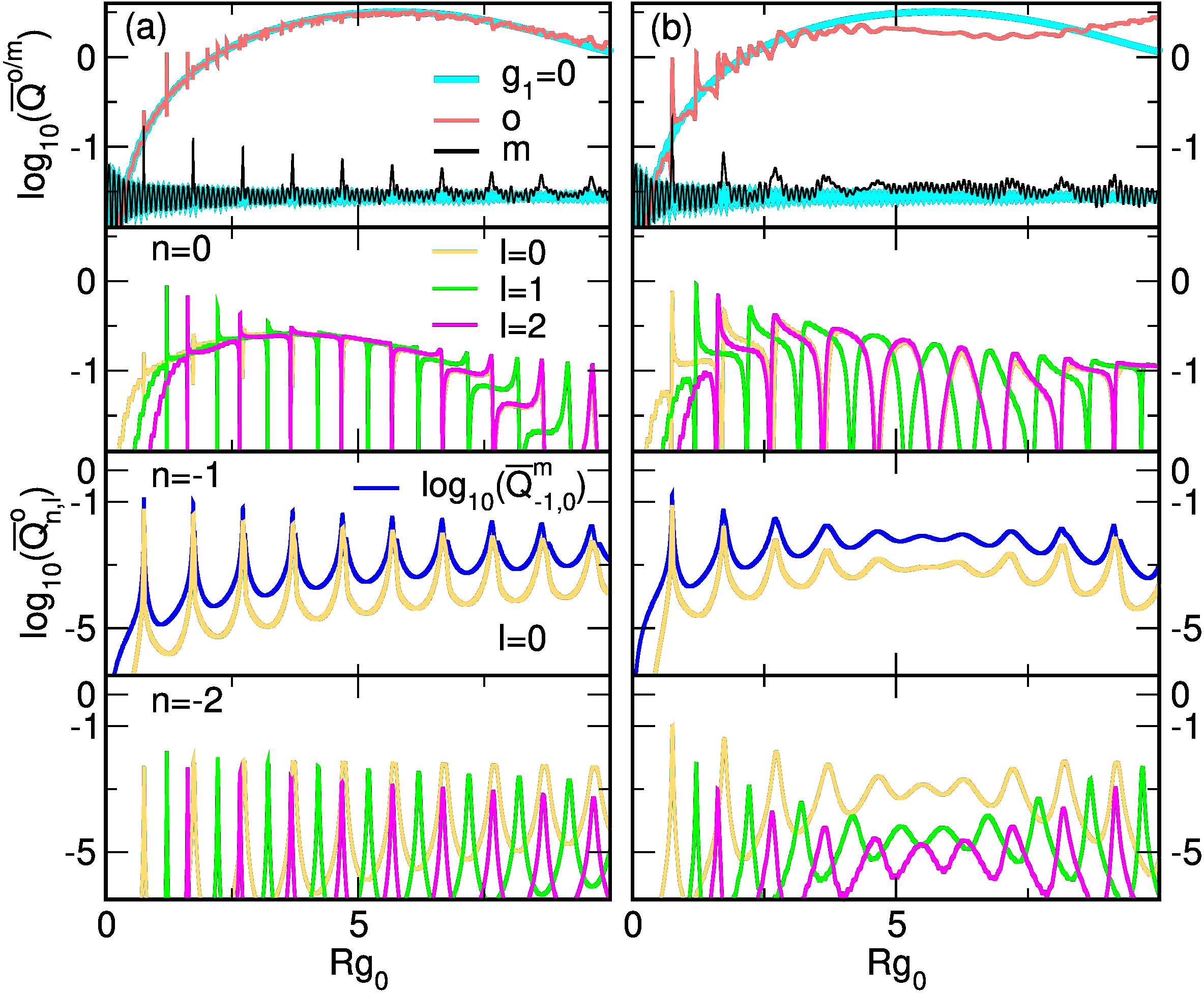}
\caption{(Color online) Time-averaged scattering efficiency (top panel) of the photon (red/gray) and the phonon (black)  and optical scattering contributions  of different partial waves (lower panels) at weak couplings, where $|g_{1}|=5\cdot 10^{-3}\Omega$  in (a) and  $|g_{1}|=0.02\Omega$ in (b). In the top panels the scattering efficiencies of the static dot are included [turquoise (thicker) line]. The scattering contribution of the phonon for the sideband $n=-1$ with $l=0$ is denoted by the blue line. To realize a symmetric Floquet resonance at $E\simeq \Omega$ [case (2) in Fig.~\ref{fig3}], we choose  $E=10^{-3}g_{0}+\Omega$, $g_{0}=\Omega\sqrt{v_{o}v_{m}}/(v_{o}+v_{m})\simeq 0.287\Omega$. \label{fig7}}
\end{figure}

Next we investigate the scattering of a photon with energy $E \simeq \Omega$, according to case (2) in Fig.~\ref{fig3}. Since the energy-coupling ratio $E/g_{0} \simeq 3.48$, the static quantum dot now acts as  a weak reflector with almost no light-sound interconversion [regime (iii) in Fig.~\ref{fig2}]. As before, the scattering by the oscillating dot is  Floquet-resonant and the situation is, in some sense,   symmetric as the energies with $n=0,-1,-2$ match the CE perfectly and the wave numbers obtained from $E_{\pm n}$ have equal magnitudes. Since $E\neq 0$ the sideband contributions are no longer symmetric with respect to $n \rightarrow -n$. 

\begin{figure}
\center
\includegraphics[width=0.42\textwidth]{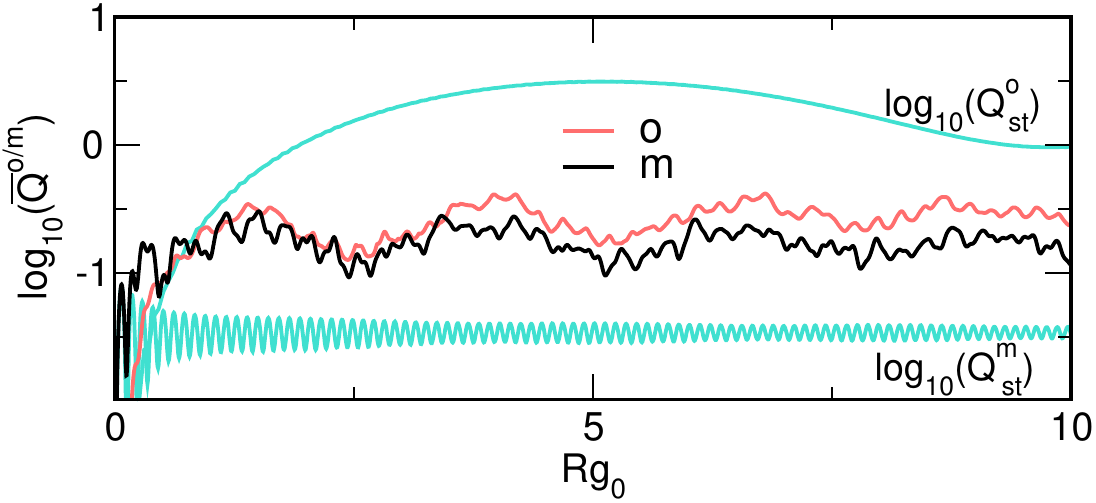}
\caption{(Color online) Time-averaged scattering efficiency of the photon  [red (gray)] and the phonon (black) slightly away from the symmetric Floquet resonance at $E=0.928 \Omega$, $g_{0}=0.3\Omega$ and $|g_{1}|=0.1\Omega$. For comparison, the corresponding scattering efficiencies of the static dot  are shown (turquoise). \label{fig8}}
\end{figure}

In Fig.~\ref{fig7} the time-averaged scattering efficiency of the photon and the phonon is depicted together with the  scattering contributions of the photon for two (weak) couplings  ($2|g_{1}|\ll \Omega$). The scattering is determined by the central band and the sidebands $n=-1,-2$; other sidebands play no role as their energies do not lie in the range of the CE, cf. Fig.~\ref{fig3}. For the mechanical mode only the $n=-1$ contribution is shown because this is the only one that modifies the scattering efficiency substantially.  Note that the size parameter $ER$ takes on large values very quickly, that is why exclusively the contributions of the first partial waves  were considered. 

While the scattering efficiency essentially follows those of the static dot, it features some very sharp resonances, see Fig.~\ref{fig7}(a). The central band contribution $n=0$ indicates that these spikes originate from resonances of the partial waves~\eqref{resonance} as they will also occur  for a static quantum dot at zero photon energy in the resonant scattering regime.  Not surprisingly, the resonant scattering regime is also reflected in the sideband contribution $n=-1$, where the effective energy-coupling ratio $E_{n=-1}/g_{0}\simeq 0$. Here, only the lowest partial wave $l=0$  is resonant, while higher partial waves are not excited due to the smallness of the effective  size parameter, $E_{n=-1}R \ll 1$. The situation changes for the sideband $n=-2$, where the effective size parameter becomes large again,  $E_{n=-2}R \gg 1$.

Increasing the coupling strength in the weak-coupling regime, the resonances broaden [compare Figs.~\ref{fig7} (b) and~(a)], and especially the low-frequency part in the functional dependence of $\overline{Q}(Rg_{0})$ markedly deviates from that of the static dot. Both effects can be attributed to larger deviations of the Floquet wave numbers from those of the static problem  when $g_{1}$ is growing. Again off-resonances occur, which becomes particularly clear for the $n=-1$ sideband contribution [see Fig.~\ref{fig7}(b)].  This signal is very similar to that one obtained in  Fig.~\ref{fig4}(b), where the same value of $g_{1}$ was used. The reason is that the effective energy-coupling ratio of the sideband is equal to that of a photon with energy at the neutrality point, $E_{n=-1}/g_{0}\simeq 0$.  This means that not only for $E\simeq 0$ but also for $E\simeq \Omega$ the interplay between  sideband and central band excitations causes a partial transition from the weak reflector regime to the resonant scattering regime  [(iii) to (i) in Fig.~\ref{fig2}], leading to the formation of a photon-dominated weak resonant scattering regime.

The scattering efficiency at moderate coupling strengths, slightly away from the symmetric Floquet resonance condition, reveals another interesting result. Figure~\ref{fig8} 
shows that in this case the scattering is no longer photon-dominated (different from  Fig.~\ref{fig7}). So while the static dot acts as a weak reflector for photons with almost no light-sound interconversion, the scattering efficiency of the phonon now  becomes comparable with that in the weak scattering regime.

\subsubsection{Floquet-resonant scattering without symmetry}

Finally, we discuss the scattering by the oscillating quantum dot for a situation without symmetry. For that we assume $E\simeq 0.12 \Omega$ and $g_{0}\simeq 0.39\Omega$, according to case (3) in Fig.~\ref{fig3}. Then the energy-coupling ratio $E/g_{0}\simeq 0.31$, and the  static dot acts as a strong reflector with angle-dependent light-sound interconversion [regime (ii) in Fig.~\ref{fig2}]. The scattering is  again Floquet-resonant.

\begin{figure}
\includegraphics[width=0.45 \textwidth]{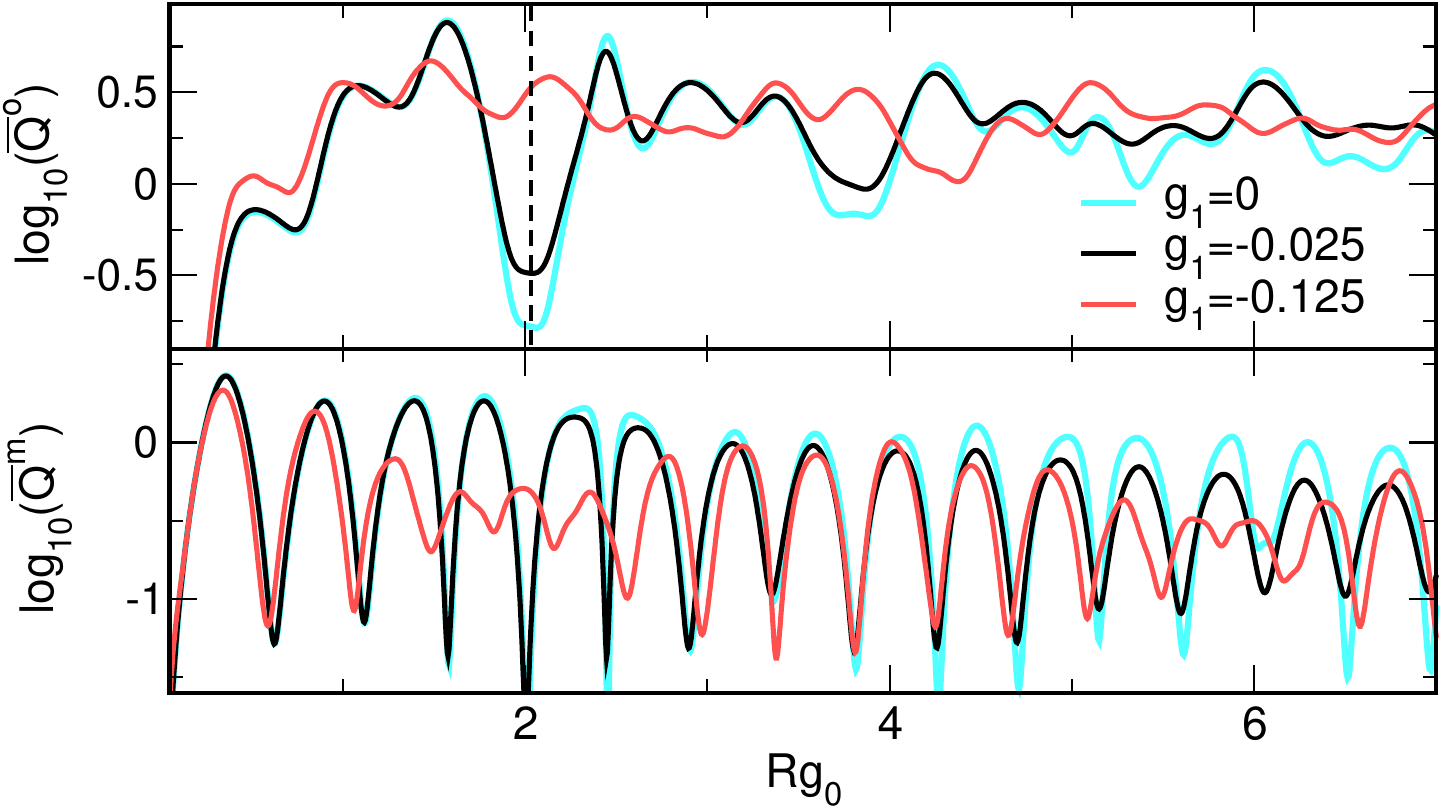}
\caption{(Color online) Time-averaged scattering efficiency at Floquet resonance without symmetry [case (3) in Fig.~\ref{fig3}]. Here $E\simeq 0.12\Omega$ and $g_{0}\simeq 0.39 \Omega$. \label{fig9} }
\end{figure}

Figure~\ref{fig9} displays the time-averaged scattering efficiency of the photon  and the phonon for  weak and  moderate coupling strength. Since the size parameter $ER\simeq 1$, the scattering efficiency of the static dot  features resonances of the first partial waves, showing up as broad peaks. The  oscillating dot weakens the resonances in the scattering efficiency of the photon as well as the light-sound interconversion rate. This effect becomes more pronounced at higher coupling strengths, and is accompanied by off-resonances  for the phonon.

\begin{figure}
\includegraphics[width=0.48 \textwidth]{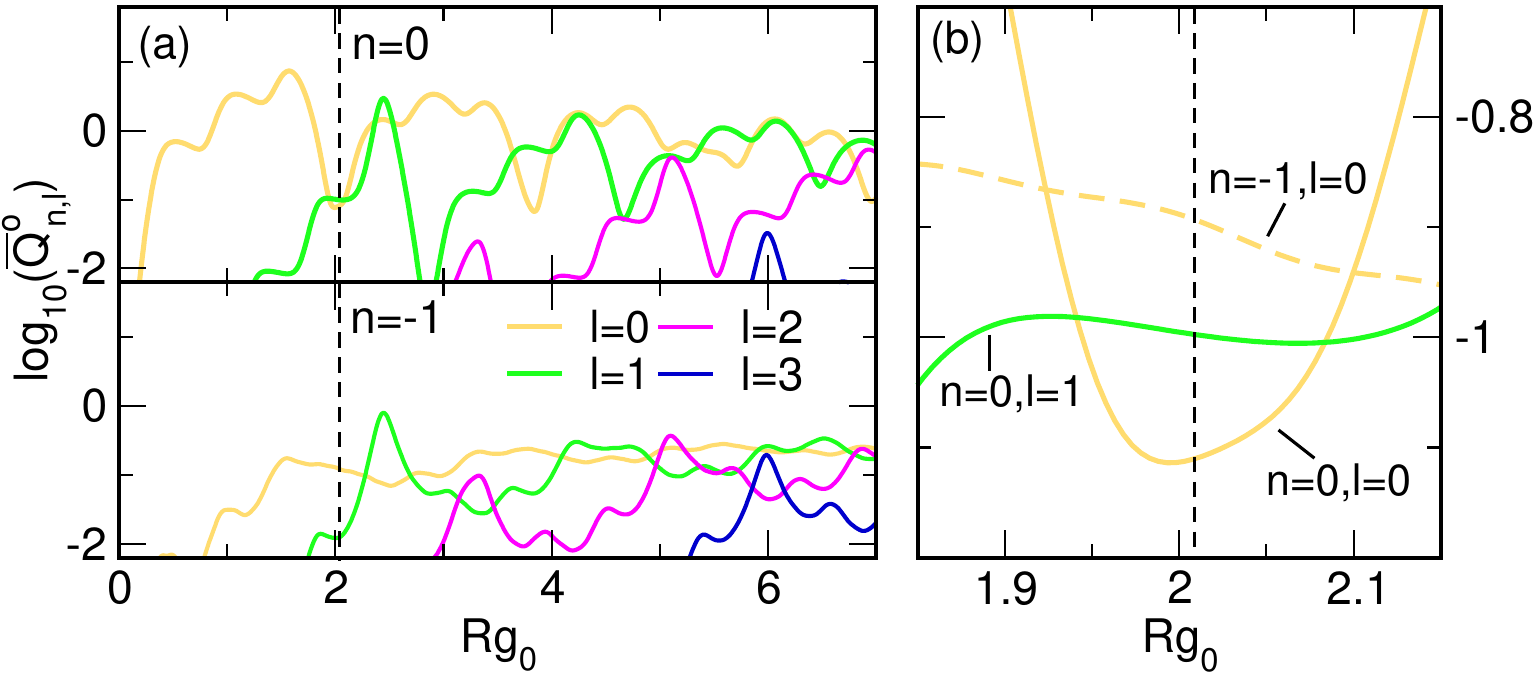}
\caption{(Color online) (a) Scattering contributions to $\overline{Q}^{o}$ given in  Fig.~\ref{fig9} for $|g_{1}|=0.025\Omega$. (b) Enlarged area near $Rg_0=2$. \label{fig10} }
\end{figure}

\begin{figure}
\includegraphics[width=0.15 \textwidth]{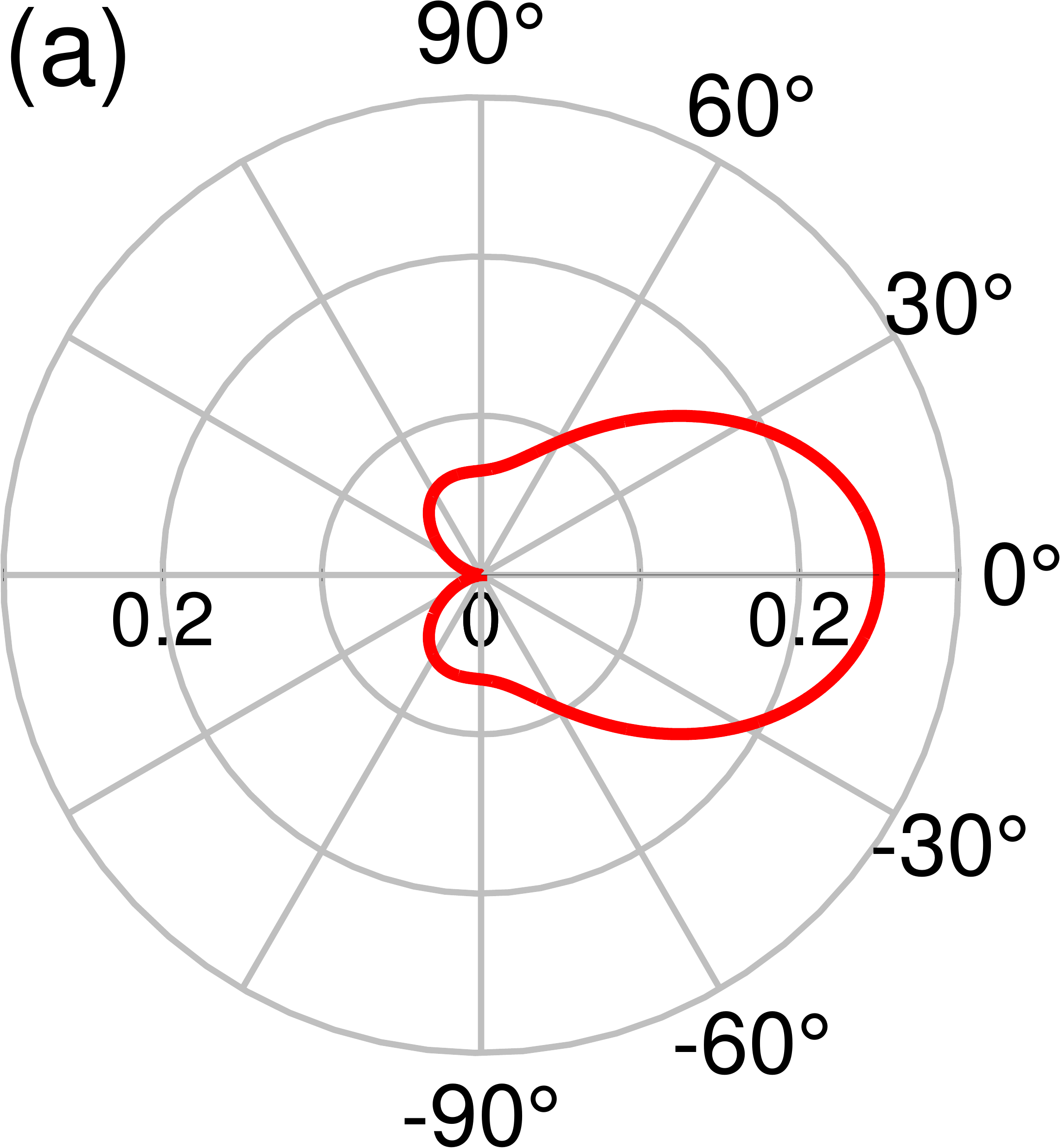}
\includegraphics[width=0.15 \textwidth]{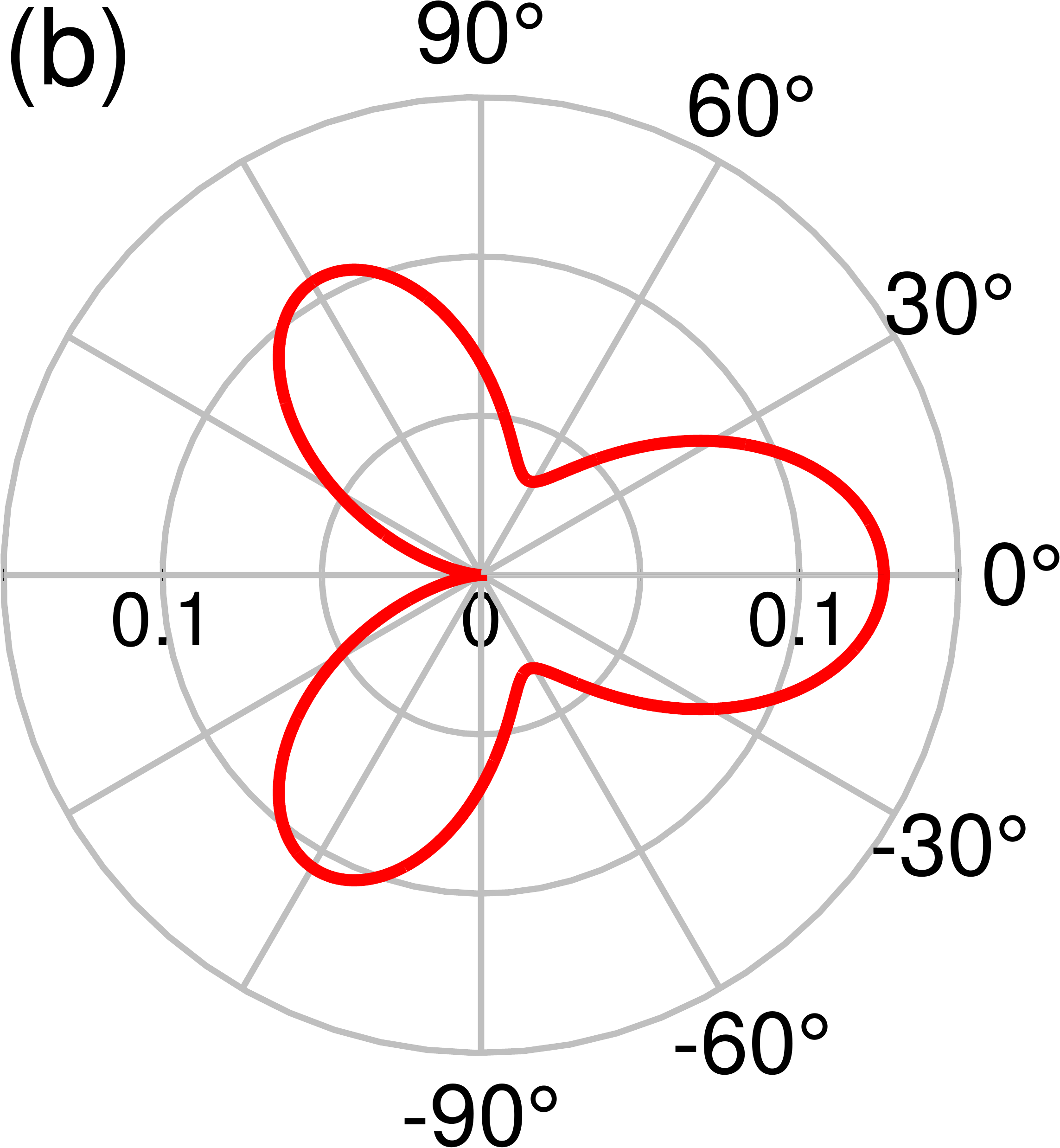}
\includegraphics[width=0.15 \textwidth]{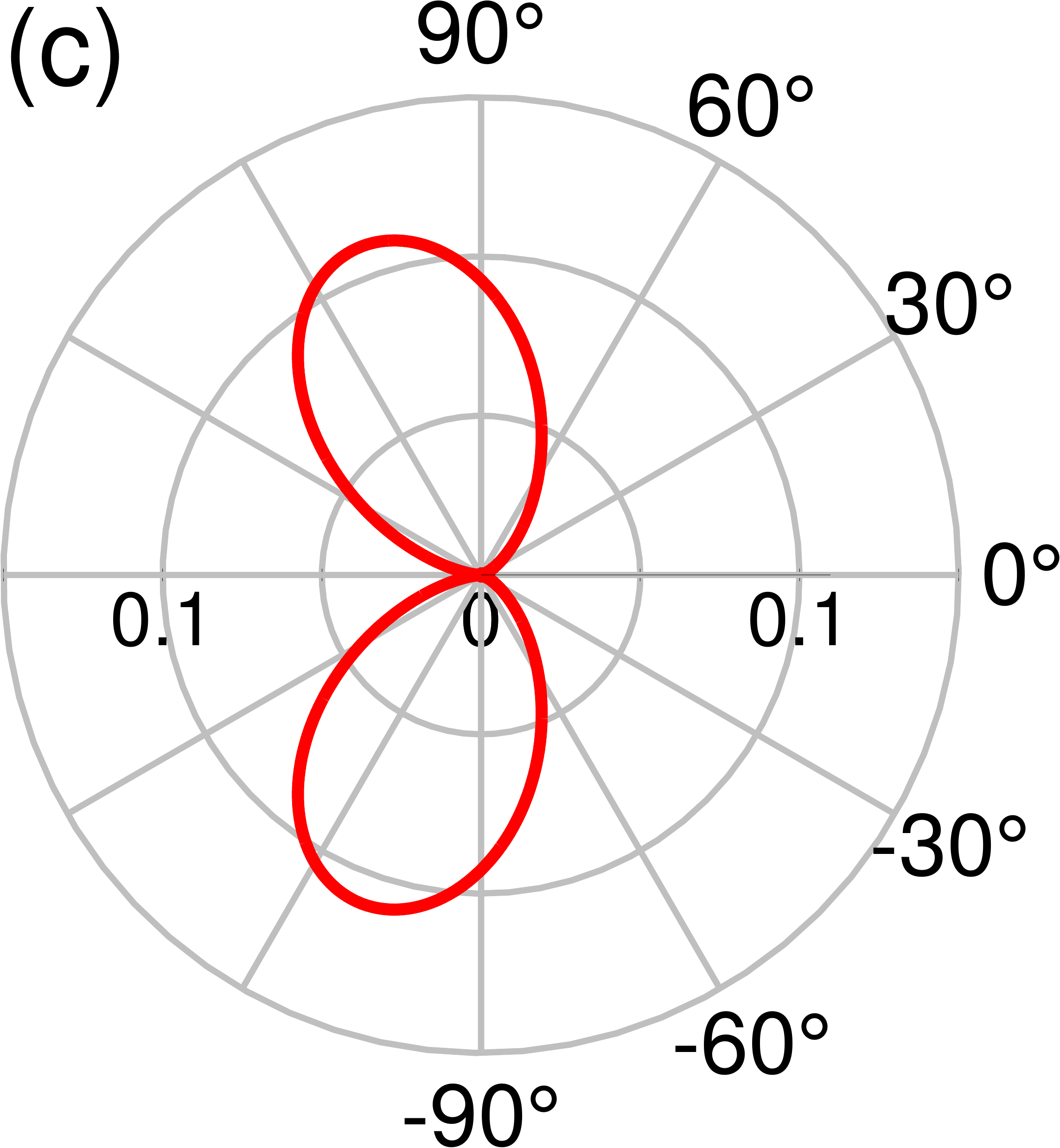}
\caption{(Color online)  Polar plot of the current density of the optical reflected wave in the far-field according to Eq.~\eqref{farfield} at the time  $t/2\pi=0.61$ (a),  $t/2\pi=1.2$ (b), and $t/2\pi=1.33$ (c) for $r=R$. Parameter values are the same as in  Fig.~\ref{fig10} with $Rg_{0}\simeq 2$ (dashed line).  \label{fig11}}
\end{figure}

Figure~\ref{fig10}~(a) gives the (relevant) photon contributions  to the scattering efficiency at weak coupling. The phononic contributions are not shown because the phonon  scattering efficiency is determined by the central band only. The sideband $n=-1$ has a significant influence on the scattering efficiency as $E_{n=-1}$ matches the CE (cf. Fig.~\ref{fig3}). Since the  corresponding effective energy-coupling ratio  $|E_{n=-1}|/g_{0} \simeq 2.3 $, the interference of states of the sideband and the central band leads to the hybridization of the weak  and the strong reflector regime  of the static dot [regimes (iii) and (ii) in Fig.~\ref{fig2}], which gives the explanation for the weakening of resonances and of the light-sound interconversion rate in Fig.~\ref{fig9}. We further observe, that only the first partial waves are excited for the sideband, although the effective size parameter is significantly larger, $|E_{n=-1}|R > E_{n=0}R$. The same effect occurs for the case of symmetric Floquet-resonant scattering at $E\simeq 0$ in  Fig.~\ref{fig4}. It seems that the size parameter $ER$ determines the maximum number of partial waves $l^{max}$ which are involved in the scattering,  whereas the  effective size parameter $E_{n \neq 0}R$ determines the maximum number of partial waves for the sidebands with the constraint $l^{max}_{n \neq 0} \leq l^{max}$ (this applies also to the Floquet scattering problem in graphene~\cite{SHF15_2}). This is reasonable, since the scattered waves with their effective size parameters merely represent the system's response, whereas the incident wave and its interaction  with the quantum dot represent the initial condition of  scattering.

Figure~\ref{fig10}~(b) enlarges the area of Fig.~\ref{fig10}(a) where the scattering contributions of different angular momentum $l$ and different energy $n$ are of comparable magnitude.  While the angular momentum defines the angle dependence of the radiation, the energy determines their time dependence [cf. Eq.~\eqref{farfield}].  
Interference has a lasting effect on the (angle- and time-dependent) radiation characteristics. This is illustrated in Fig.~\ref{fig11}.  At different points in time the interference causes either (a) forward scattering due to the $l=0$ mode, (b) scattering in several directions due to the $l=1$ mode, or (c) the absence of forward scattering (Fano resonance) due to the interference of the $l=0$ and $l=1$ modes~\cite{WF17}.  In this way, the oscillating quantum dot might act as a time-dependent photon transistor.

\section{Conclusions}\label{conc}
The main goal of this work was to examine the time-dependent scattering of  two-fold degenerate Dirac-Weyl quasiparticles by laser-driven quantum dots in optomechanical graphene. The setup considered models the propagation and interconversion of light and sound on a honeycomb array of optomechanical cells, structured by circular, oscillating 
(photon-phonon-coupling) barriers.   

As our investigations have shown, the temporal modulation ($\Omega$) of the photon-phonon coupling in the quantum dot region ($R$) tremendously influences the quasiparticle transport. Here,  unlike  the energy-conserving case of a static quantum dot where the scattering is essentially determined by the ratio between the energy of the incident photon wave and the coupling strength of the barrier, inelastic scattering  gives rise to the excitation of sideband states with energies $E_{n}=E+n\hbar \Omega$. Their interference causes a mixing  of long-wavelength (quantum) and short-wavelength (quasiclassical) regimes. The number of sidebands involved is greater the larger (smaller) the amplitude (frequency) of the barrier oscillation. This affects also the effective size parameters $E_{n}R$, which determine the angular momentum contributions involved in the scattering process.  The consequence is a time-periodic, strongly angle-dependent emission of light and sound (with Fano resonances),  analogous to  electron transport through driven graphene quantum dots. In this way, the optomechanical quantum dot acts as a time-dependent converter for photons and phonons. 

Analyzing the underlying, effective two-level system within Floquet theory, it was shown that avoided crossings in the quasienergy band structure  are of particular importance. More specifically, when the (sideband) energy lies in the vicinity of an avoided crossing (Floquet resonance), the influence of the barrier is most prominent since  the wave numbers determining the scattering process most deviate from those of the static dot. Then even a small oscillation amplitude may significantly affect the scattering, up to the point where the light-sound interconversion is suppressed and revived in the course of interference of waves with different wave numbers. 

The results presented in this work should have impact on both, fundamental problems such as the observation of \textit{zitterbewegung} and  potential applications based on quantum-optical, laser-driven optomechanical metamaterials being suitable for the transport, storage, and transduction of photons and phonons.  In this context, a more realistic description of optomechanical systems beyond the continuum approximation, which ideally  involves wave-packet dynamics and dissipation, is highly desirable, as well as more in-depth studies about the role  of time-dependent (synthetically generated) magnetic fields~\cite{PBSM15}.

\begin{acknowledgments}
The authors would like to thank K. Rasek for valuable discussions.
\end{acknowledgments}

\section*{APPENDIX A: IMPLEMENTATION OF THE FLOQUET APPROACH}\label{app}

Inserting the Floquet state~\eqref{FS} into the time-dependent Dirac equation yields the Floquet eigenvalue equation: 
\begin{eqnarray}\label{FEE}
&&\sum \limits_{p}\sum\limits_{\tau=\pm} \Big\{c_{p}^{\tau}\bk{E^{\tau,\sigma}+p\Omega}\ket{\tau}\delta_{pp'} \nonumber \\ 
&&+g_{1}\sum \limits_{\tau '=\pm}c_{p}^{\tau}\alpha^{\tau}_{\tau '}\ket{\tau'} \bk{\delta_{p+1,p'}+\delta_{p-1,p'}}\Big\}
\nonumber \\
&&=\varepsilon\sum \limits_{p} \sum\limits_{\tau=\pm} c_{p}^{\tau}\ket{\tau}\delta_{pp'},\quad  p'\in \mathbb{Z},
\end{eqnarray}
where 
\begin{equation}\label{static_disp}
E^{\tau}=\overline{v}\sigma q+\sigma \tau \sqrt{g_{0}^2+\delta v ^2 q^2/4}
\end{equation} 
is the energy dispersion of the time-independent problem for wave number $q$, and
\begin{equation}\label{Fcoeff}
\alpha^{\tau}_{+}=\frac{\mathcal{N^{\tau}}}{\mathcal{N^{+}}}\frac{g_{0}^{2}-\gamma^{\tau}\gamma^{+}}{g_{0}\bk{\gamma^{+}-\gamma^{-}}}=-\tau \alpha^{-\tau}_{-},
\end{equation} 
with the normalization factor,
\begin{equation}
\mathcal{N^{\tau}}=1/\sqrt{g_{0}^2+(\gamma^{\tau})^2}, \quad \gamma^{\tau}=v_{o}\sigma q-E^{\tau}.
\end{equation}
 Based on Eq.~\eqref{FEE} we define the vector of Fourier components, $\bvec{c}=(\ldots,c_{-1}^{+},c_{-1}^{-},c_{0}^{+},c_{0}^{-},c_{1}^{+},c_{1}^{-},\ldots)^{\text{T}}$, and the (Hermitian) Floquet matrix,
\begin{equation}\label{app:FEE}
\mathcal{F}=\begin{pmatrix}
\ddots &  &  &  & &  &  & \\
 & E^{+}_{-1} & 0 & g_{1}\alpha^{+}_{+} & g_{1}\alpha^{-}_{+} & 0 & 0 & \\
 & 0 & E^{-}_{-1} & g_{1}\alpha^{+}_{-} & g_{1}\alpha^{-}_{-} & 0 & 0 & \\
 &  g_{1}\alpha^{+}_{+} &  g_{1}\alpha^{-}_{+} & E^{+} & 0 & g_{1}\alpha^{+}_{+} &  g_{1}\alpha^{-}_{+} & \\
 &  g_{1}\alpha^{+}_{-} &  g_{1}\alpha^{-}_{-} & 0 & E^{-} & g_{1}\alpha^{+}_{-} & g_{1}\alpha^{-}_{-}  & \\
 & 0 & 0 & g_{1}\alpha^{+}_{+} &  g_{1}\alpha^{-}_{+}  & E^{+}_{+1}& 0 & \\
 & 0 & 0 &  g_{1}\alpha^{+}_{-} &  g_{1}\alpha^{-}_{-} & 0 & E^{-}_{+1} & \\
 &  &  &  &  &  &  & \ddots
\end{pmatrix},
\end{equation} 
for $E_{n}^{\tau}=E^{\tau}+ n \Omega$. 
\begin{figure}
\center
\includegraphics[width=0.48 \textwidth]{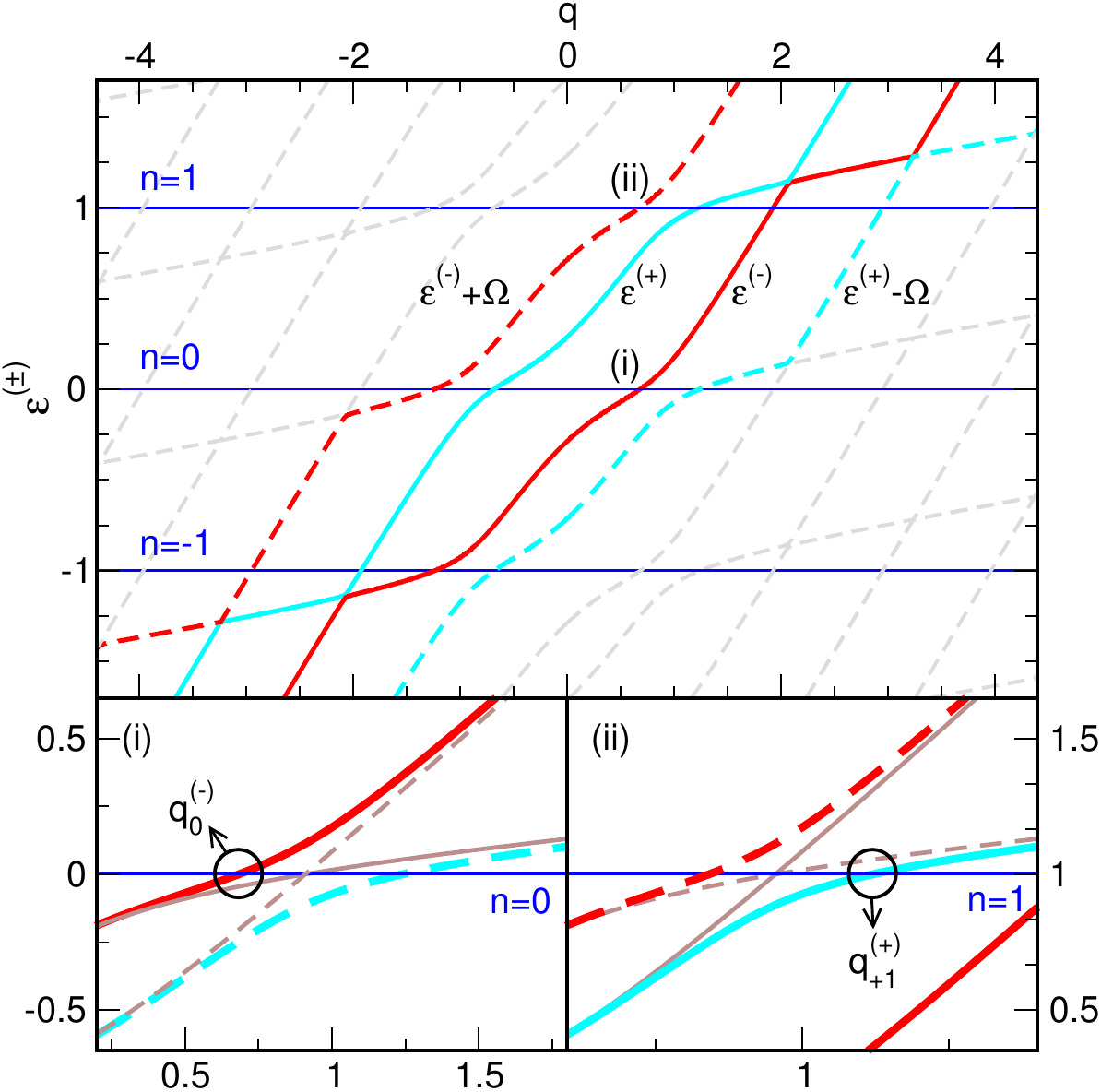}
\caption{(Color online) Quasienergies $\varepsilon^{\bk{\pm}}+ n\Omega$, $n \in \mathbb{Z}$, obtained as eigenvalues of the Floquet matrix~\eqref{app:FEE} for  $|g_{1}|=0.14 \Omega$ as a function of the wave number $q$ (using  $\sigma=\pm 1$ for positive and negative values of $q$).   Inserted are also the  energies $E_{n}=E+n\Omega$ of the central band $n=0$ and the sidebands $n=\pm 1$ (blue horizontal lines) for $E=0$ and  static coupling $g_{0}=0.287 \Omega$, corresponding to the case of symmetric Floquet resonance close by $E\simeq 0$ (see discussion in Sec.~\ref{symmetric} of the main text). The proper pair of quasienergies $\varepsilon^{\bk{\pm}}$ (solid lines) that has to be used for the scattering problem is that which coincides with the dispersion of the static case at $q\rightarrow 0$. The wave numbers used for the scattering problem are determined by the zeros of $E_{n}-\varepsilon^{\bk{\pm}}\bk{q}$ and are marked exemplary for the cases $n=0, 1$ in the lower panels (i) and (ii). For comparison the polariton branches of the energy dispersion of the static case, $E^{\mp}\bk{q}$ (solid) and $E^{\pm}\bk{q}\mp  \Omega$ (dashed), are shown (brown thin lines). Since $E=0$, the wave numbers reveal the symmetry $q^{\bk{\pm}}_{-n}=- q^{\bk{\mp}}_{n}$.} \label{fig12}
\end{figure}
We fix $\Omega=1$, which is justified due to the scale invariance of the scattering problem.  The quasienergies $\varepsilon$  in  $\mathcal{F}\bvec{c}=\varepsilon \bvec{c}$ are obtained as the eigenvalues of  the Floquet matrix~\eqref{app:FEE} and depend on the two barrier parameters $g_{0},g_{1}$, as well as on  wave number $q$. The pseudospin projection $\sigma=\pm 1$ leads only to a change in the sign of the quasienergies  and is determined by the sign of the wave number. As a consequence of the polariton degree of freedom  $\tau=\pm 1$, the static dispersion~\eqref{static_disp} is two fold degenerate. Accordingly, the quasienergies are two fold degenerate, too, which is reflected in the block-diagonal form of $\mathcal{F}$ and is marked by the index $\bk{\pm}$ hereinafter. Diagonalization yields a pair of  quasienergies $\varepsilon^{\bk{\pm}}\bk{q}$ with Fourier vectors $\bvec{c}^{\bk{\pm}}\bk{q}$ for each $q$. Other pairs of quasienergies $\varepsilon^{\bk{\pm}}\bk{q}+n\Omega$ are also eigensolutions of Eq.~\eqref{app:FEE}, but in principle they all contain the same information about the time dependence. 

According to Eq.~\eqref{trans}, the eigensolutions of $\mathcal{F}$ are needed to construct the transmitted wave function inside the dot. Since the oscillating barrier shifts the energy $E$ of the incoming wave, $E_{n}=E+n \Omega$,  the quasienergies are fixed: $\varepsilon^{(\pm)}(q)=E_{n}$. The zeros of $\varepsilon^{(\pm)}(q)-E_{n}$ yield the wave numbers $q^{\bk{\pm}}_{n}$, and hence the Fourier vectors $\bvec{c}^{\bk{\pm}}_{n}$ can be calculated. Doing this  it makes sense to connect the considered  pair  of quasienergies  with the energy dispersion in the static case for $q \rightarrow 0$:  $\varepsilon^{\bk{\pm}}\bk{q\rightarrow 0}=E^{\pm}\bk{q\rightarrow 0}$. We note that when using the Floquet approach the specific geometry of the barrier only enters the scattering matrix via Eqs.~\eqref{scatmatrix1} and~\eqref{scatmatrix2} (e.g., the results for a  planar barrier are given in~\citep{WF18}).   

Figure~\ref{fig12} displays the highly symmetric situation that evolves in the numerical work  for the Floquet resonance at photon energy $E\simeq 0$ discussed in the main text. By tracking the quasienergies in dependence of $q$, the condition $\varepsilon^{\bk{\pm}}(q)=E_{n}$ defines the wave numbers $q_{n}^{\bk{\pm}}$ (and Fourier vectors) that have to be used for the barrier wave function (crossings of the blue horizontal lines with the quasienergies); see panels (i) and (ii) for  $n=0,1$. Deviations of the wave numbers $q$ from those of the dispersion of the static case (obtained from crossings of the horizontal lines with the brown thin lines in lower panels of Fig.~\ref{fig12}) arise due to the avoided crossings. Obviously, these deviations are largest in the vicinity of the points where the two polariton branches of the static dispersion cross each other.  The corresponding crossing energies  are given by Eq.~\eqref{crossings}. Of course, the influence of the oscillating barrier on the scattering is most prominent for energies $E_{n}$ near a crossing energy. There even small couplings $|g_{1}| \ll \Omega$  significantly modify the scattering (cf. Figs.~\ref{fig3},~\ref{fig6}, and~\ref{fig8}.

We finally note that at larger $|q|$ values the quasienergies are less affected by the barrier; for $|q| \gg 1$ the quasienergy and the dispersion of free quasiparticles merge. This can be used to implement  truncation criteria for the number of sidebands $n_{max}$ which will have to be considered in the numerical work. Taking into account that $2|g_{1}|\le g_{0}$, we found that $\text{dim}\mathcal{F}\simeq 2+4(x-1)$ with $x=2(1+10\cdot 4|g_{1}|)$ serves as a good estimate for numerical convergence of the quasienergies $\varepsilon^{\bk{\pm}}$ as well as for those of  the scattering coefficients. Then the  maximum number of sidebands used in the numerics  should be at least $n_{max}=x/2$, i.e.,  $n_{max}\simeq 1+10\cdot 4|g_{1}|$. 

\bibliographystyle{apsrev4-1}
%


\end{document}